\documentclass[letterpaper,twocolumn,screen,10pt]{article}
\usepackage{usenix}

\usepackage{xspace} 
\usepackage{enumitem}
\usepackage{color}
\usepackage{colortbl}
\usepackage{xcolor}
\usepackage{pifont}

\usepackage{epstopdf}
\usepackage{graphicx}
\usepackage[labelfont=bf]{caption}
\usepackage{subcaption}
\usepackage{pgfplots}
\usepackage{pgfplotstable}

\usepackage[capitalise]{cleveref}

\usepackage{siunitx}
\usepackage{ifthen}
\usepackage{bm}

\usepackage{listings}
\usepackage[utf8]{inputenc}

\usepackage{booktabs}  
\usepackage{tabu}
\usepackage{array}
\usepackage{multirow}
\usepackage{diagbox}

\usepackage{balance}
\usepackage{microtype}

\usepackage[english]{babel}
\usepackage{blindtext}

\def\setuppreprint{0}
\def\setupcameready{1}

\newcommand{\name}{\textsc{OptiReduce}\xspace}

\newcommand{\titleinfo}{\Large \name{}: Resilient and Tail-Optimal AllReduce for \\Distributed Deep Learning in the Cloud} 
\newcommand{\authorinfo}{\em Ertza Warraich, Omer Shabtai$^{\dagger}$, Khalid Manaa$^{\dagger}$, Shay Vargaftik$^{\ddagger}$, \\\em Yonatan Piasetzky$^{\dagger}$, Matty Kadosh$^{\dagger}$, Lalith Suresh$^{\S}$, Muhammad Shahbaz$^{*}$}
\newcommand{\institutioninfo}{Purdue University~~~$^{\dagger}$Nvidia~~~$^{\ddagger}$VMware Research~~~$^{\S}$Feldera~~~
$^{*}$University of Michigan}
\newcommand{\submissioninfo}{\em Submission \#1, \pageref{lastpage} Pages Body, \pageref{totalpage} Pages Total}

\if\setupcameready1
	\def\shownames{1} 										
	\def\showpagenumbers{0}                                 
    \def\showcomments{0} 									
    \def\showacks{1} 										
\else
	\def\shownames{0}										
	\def\showpagenumbers{1}                                 
    \if\setuppreprint1										
        \def\showcomments{0}
    \else
        \def\showcomments{1}
    \fi
    \def\showacks{0}										
\fi

\newcommand{\ie}{i.e.}
\newcommand{\eg}{e.g.}

\setlist[itemize]{topsep=4pt, itemsep=4pt, parsep=1.5pt}

\if\showcomments1
    \usepackage[colorinlistoftodos,textsize=tiny]{todonotes}
\else
    \usepackage[disable]{todonotes}
\fi

\begin{document}

\title{\titleinfo}

\if\shownames1
    \author{\authorinfo \\ \institutioninfo}
\else
    \author{\submissioninfo}
\fi

\if\showcomments1
	\onecolumn
    \setcounter{page}{0}
    \listoftodos{}
    \clearpage
    \twocolumn
    \setcounter{page}{1}
\fi

\if\showpagenumbers0
	\pagestyle{empty}
\fi

\maketitle

\begin{abstract}
We present \name{}, a new collective-communication system for the cloud with bounded, predictable completion times for deep-learning jobs in the presence of varying computation (stragglers) and communication (congestion and gradient drops) variabilities.  
\name{} exploits the inherent resiliency and the stochastic nature of distributed deep-learning (DDL) training and fine-tuning to work with approximated (or lost) gradients---providing an efficient balance between tail performance and the resulting accuracy of the trained models.

Exploiting this domain-specific characteristic of DDL, \name{} introduces (1) mechanisms (\eg, unreliable bounded transport with adaptive timeout) to improve the DDL jobs' tail execution time, and (2) strategies (\eg, Transpose AllReduce and Hadamard Transform) to mitigate the impact of dropped gradient entries on model accuracy.
Our evaluation shows that \name{} achieves 70\% and 30\% faster time-to-accuracy (TTA), on average, when operating in shared, cloud environments (\eg, CloudLab) compared to Gloo and NCCL, respectively.
\end{abstract}

\vspace{-2pt}
\section{Introduction}
\label{sec:introduction}

Synchronous distributed data-parallel training~\cite{zinkevich2010parallelized} is now the de-facto standard for training and fine-tuning large-scale deep-learning models (comprising billions or even trillions of parameters) and datasets (comprising terabytes of data) that form the backbone of many mainstream enterprise applications, including computer vision~\cite{du2017fused, goel2020survey, koziarski2017image, xu2019innohar}, natural-language processing and large-language models~\cite{yin2017comparative, liu2019multi, devlin2018bert,brown2020language}, recommendation and prediction systems~\cite{huang2019trec, fu2018novel, ramesh2021optimized, feng2019using, heredia2018social}, and healthcare~\cite{qi2020image, khan2020coronet, mallick2019brain, yuan2019deep}.
Under this scheme, the training occurs in rounds (or epochs). 
Workers locally train a copy of the model on a fragment of data and then share the model updates (\ie, gradients) among themselves over the network to compute an aggregated result. 
The aggregate is then used to update the model locally for the next round of training. 
Distributed deep-learning (DDL) is, therefore, inherently a computation- and communication-intensive workload and is becoming even more so with the growing model sizes (\eg, Bart~\cite{lewis2019bart}, GPT-2/3~\cite{radford2019language,brown2020language}, LLaMA~\cite{touvron2023llama,dubey2024llama}), and datasets~\cite{russakovsky2015imagenet,sun2017revisiting,criteo}.

To train and fine-tune such large models, extensive efforts are underway in reducing both the computation and communication time of DDL jobs, albeit in isolation.
On the one hand, we have GPUs~\cite{steinkraus2005using} and emerging hardware accelerators, like Tensor Processing Units (TPUs)~\cite{jouppi2017datacenter}, that are drastically bringing down the computation time---reducing it by 62$\times$ over the last decade~\cite{sapio2021scaling}.
While, on the other hand, we have recent proposals based on programmable switches~\cite{xavier2021programmable} that aim at reducing the communication time by 2--5$\times$ (via in-network aggregation)~\cite{sapio2021scaling}.
Yet, when seen together, both these efforts mainly help in improving the average completion time of a deep-learning job (either by accelerating computation or communication). 
The vast array of system-level variabilities (\eg, device failures, OS and hypervisor scheduling, and resource contention) and network-level delays (\eg, congestion and retransmissions) still lead to long tails; hence, resulting in poor overall performance for these training jobs---with tail reaching as high as 4$\times$ the mean latency in shared environments (\eg, public cloud providers)\footnote{Cloud providers typically do not offer preferential treatment to small tenants, but even large tenants with dedicated racks face long tails when communicating across racks in the provider's network. Private communication with a hyperscaler.}~\cite{wang2024towards,guo2016rdma,narayanan2019pipedream,li2020pytorch,lao2021atp,yang2022using,gangidi2024rdma}.

\begin{figure*}
     \centering        
     \includegraphics[width=0.92\linewidth]{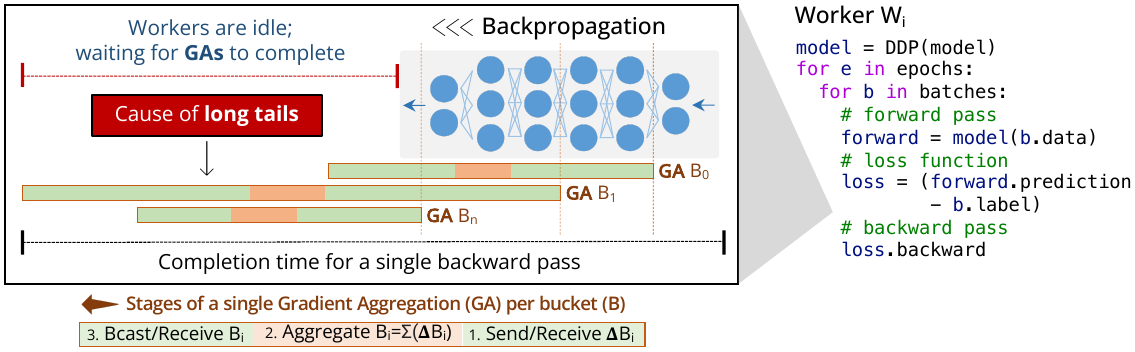}
     \vspace{-4pt}
     \caption{\bf A backpropagation pass in distributed data-parallel (DDP) training. Multiple gradient aggregation (GA) runs share a bucket ($B_i$) worth of gradient entries among worker nodes ($W_n$), in parallel. The two send(bcast)/receive stages (1, 3) in GA incur the most time---contributing to the tail latency and stalling workers.}
     \label{fig:epochs}
     \vspace{-17pt}
\end{figure*}

In this paper, we make the case for \name{}, a collective-communication system for the cloud tenants that ensures bounded, predictable completion times for deep-learning jobs in the presence of myriad computation and communication variabilities. 
Public clouds are becoming increasingly appealing for training, and more specifically fine-tuning, large foundation models~\cite{bommasani2021opportunities}, for enterprises and individuals lacking resources to set their own in-house distributed training clusters~\cite{gvr, runai, redresscompliance, hyperstack, mentormate, orange-business, forbes, encapture, symphony-solutions}.
\name{} exploits the inherent resiliency and the stochastic nature of deep-learning systems to work with approximated or lost gradients and provides an efficient balance between tail performance and the resulting accuracy of the trained models.
Others are already utilizing this characteristic of deep learning to optimize DDL hardware design (\eg, chip area~\cite{10.1145/3394885.3431632, rucker2021chopping}), minimize traffic overhead~\cite{lin2017deep, wangni2017gradient, fei2021efficient, li2024thc}, or offload certain DDL tasks to the network switches~\cite{xiong2019switches, sapio2021scaling, xavier2021programmable, lao2021atp, wang2024towards}.
For instance, to reduce communication time, ATP~\cite{lao2021atp} and SwitchML~\cite{sapio2021scaling} leverage fixed-point arithmetic for gradient aggregation in programmable switches with acceptable approximation loss, whereas MLT~\cite{wang2024towards} prioritizes and drops packets inside switches based on model layers and gradient magnitudes to limit loss in accuracy.
Various gradient-compression schemes~\cite{lin2017deep, wangni2017gradient, fei2021efficient, li2024thc} employ lossy compression to reduce network traffic overhead (\eg, total bytes transferred) while limiting deviation from the achievable model accuracy.
Similarly, hardware designers are incorporating approximate operations (\eg, approx. multipliers~\cite{10.1145/3394885.3431632, rucker2021chopping}) in their architectures to minimize resource and energy usage---to scale to ever-increasing DDL models.
However, these solutions are still susceptible to tail effects (\eg, slow workers and network variabilities)~\cite{alizadeh2010data,raiciu2011improving,zats2012detail,dean2013tail,guo2015pingmesh,zhang2016treadmill,li2016lossradar,sriraman2017deconstructing, wang2024towards}, and are not optimized for cloud environments, often times requiring direct access to the provider's network infrastructure.

In \name{}, we exploit this resiliency and replace the (tail-prone) deterministic, run-to-completion stages of an AllReduce collective in DDL, with best-effort, \emph{time-bounded} implementations.
\begin{itemize}[leftmargin=*]
	\item \name{} introduces a {\em Transpose-Allreduce Collective (TAR)} to reduce the impact of lost gradient entries by establishing direct peer-to-peer communication among nodes in each round, rather than propagating the entries through all nodes as in Ring~\cite{patarasuk2009bandwidth}.\vspace{-3pt}
	\item It also implements an {\em Unreliable Bounded Transport (UBT)} to maximize the number of gradient entries received during each window. 
UBT introduces the notion of {\em adaptive timeout} to restrict the time a deep-learning job spends doing computation (aggregation) and communication (gradient sharing). 
Furthermore, it adds support for {\em dynamic incast} to dynamically adjust the number of concurrent senders per receiver in each round, thus optimizing communication by reducing the total rounds required for gradient aggregation.\vspace{-3pt}
	\item Lastly, to minimize the impact of missed or dropped gradient entries, \name{} employs the {\em Hadamard Transform (HT)}~\cite{pratt1969hadamard} to ensure, for any drop pattern (\eg, tail drops), a receiver still obtains an unbiased estimate of the model's gradients resulting in a minimal loss in accuracy.
\end{itemize}

We implement \name{} as a new AllReduce scheme inside Gloo~\cite{gloo},\footnote{We pick Gloo for its ability to use both GPUs and CPUs (more suitable for a cloud environment), but \name{} can operate with other libraries as well (\eg, NCCL~\cite{jeaugey2017nccl} or MSCCL~\cite{msccl}).} a popular collective-communication library.
Doing so makes \name{} immediately compatible with existing DDL frameworks (\eg, PyTorch) without modifications.
We run our experiments on various popular large deep-learning models (including BART~\cite{lewis2019bart}, OpenAI's GPT-2~\cite{radford2019language}, and Meta's Llama 3.2) and evaluate \name{} on CloudLab~\cite{duplyakin2019design}---a public cloud facility for researchers---as well as under different shared environment settings using a local virtualized cluster, with varying tail-to-median latency ratios.
We have made our complete \name{} prototype~\cite{ultima-prototype} publicly available at \href{https://optireduce.github.io/}{https://optireduce.github.io}.

Our evaluation demonstrates that \name{} achieves, on average, 57\% and 25\% faster time-to-accuracy (TTA) on CloudLab compared to Gloo~\cite{gloo} and NCCL~\cite{jeaugey2017nccl}, respectively. 
These performance gains are even more pronounced in environments with larger tail-to-median latency differences, where \name{} outperforms Gloo by 91\% and NCCL by 35\% (\S\ref{sec:evaluation}).
We observe that it is the latency of the gradient aggregation (GA) step (\Cref{fig:epochs}) that inflates three folds when operating under tail-heavy environments (\eg, public clouds), hence strengthening the need for a new collective like \name{}.
We also perform a deeper analysis of the various components of \name{}.
For example, in our evaluation, enabling Hadamard Transform mitigates the impact of tail-drops and improves TTA by 1.8$\times$ even when up to 10\% of gradient entries are lost.

We begin with a background on distributed deep-learning (DDL) training---more specifically, distributed data-parallel (DDP) training---and the impact of stragglers on performance (\S\ref{sec:background}). 
We then make a case for and present a design (\S\ref{sec:design}) and implementation (\S\ref{sec:implementation}) of a new communication-collective system, \name{}, and evaluate how it exploits DDL's resiliency against gradient loss to improve performance (\S\ref{sec:evaluation}).

\vspace{-2pt}
\section{Background \& Motivation}
\label{sec:background}

\vspace{-2pt}
\subsection{Distributed Deep Learning \& Stragglers}
\label{ssec:ddl-stragglers}
Distributed deep learning (DDL) helps scale (and speedup) model training by utilizing an increasing number of hardware accelerators, \eg, GPUs~\cite{steinkraus2005using} and TPUs~\cite{jouppi2017datacenter}, across server nodes~\cite{abadi2016tensorflow,chen2015mxnet,moritz2015sparknet}.
To do so, it employs two approaches: {\em (i)} distributed data parallelism (DDP)~\cite{zinkevich2010parallelized} to run batches\footnote{A batch is a set of training data used in a given forward/backward pass.} on multiple accelerators in parallel, and {\em (ii)} distributed model parallelism (DMP)~\cite{karakus2021amazon} to deploy larger models that fail to fit on a single accelerator.
These approaches are orthogonal and can be used in conjunction. 
We focus on DDP in this paper to highlight our contributions (\S\ref{sec:design}).

In distributed data parallelism (DDP), multiple worker nodes run the same deep-learning model on their portion of a training dataset, distributed evenly across nodes.
Each portion is further subdivided into batches, which are processed sequentially (\ie, forward pass, loss function, and backward pass) in each epoch, \Cref{fig:epochs}.
During the forward pass, the model (\eg, a neural network) operates on a batch and generates a prediction, which is then compared with the ground truth (\eg, label) to calculate the model's loss.
Next, the backward pass computes gradients using a loss function, which is used by an optimization algorithm (\eg, Stochastic Gradient Descent, SGD~\cite{kiefer1952stochastic,bottou2012stochastic}) to update the model parameters.
Finally, to ensure all workers learn from what others have learned from their portion of the dataset, these gradients are averaged (reduced) and shared across all nodes, after each backward pass.

The process of accumulating, reducing, and sharing gradients back with worker nodes is referred to as {\em gradient aggregation} (or reduction)~\cite{dean2012large}.
Originally, reduction used to happen strictly after the backward pass; however, more recently, to hide communication latency, modern frameworks (like PyTorch~\cite{li2020pytorch}) overlap it with the backward pass (\Cref{fig:epochs}).
As soon as a bucket ($B$) worth of gradient entries becomes available on a node, it is sent for reduction.\footnote{PyTorch limits these simultaneous reduction operations to two~\cite{li2020pytorch}.}

\begin{figure}
     \centering        
     \includegraphics[width=0.88\linewidth]{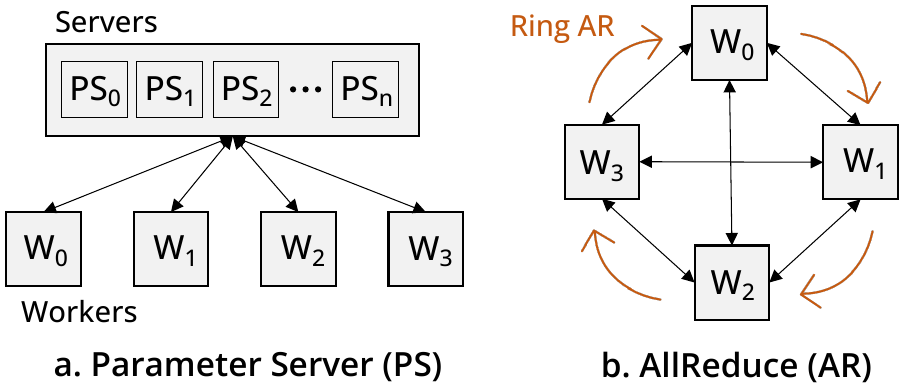}
     \vspace{-10pt}
     \caption{\bf Architectures for gradient aggregation: Parameter Server (PS) and AllReduce (AR).}
     \label{fig:reduce-arch}
     \vspace{-18pt}
\end{figure}

Two common architectures for gradient aggregation are: Parameter Server (PS)~\cite{li2013parameter} and AllReduce (AR)~\cite{clarke1994mpi}. 
In the PS architecture (\Cref{fig:reduce-arch}a), a central server (or group of servers) receives (gathers) gradients from participating workers, aggregates (reduces) them, and broadcasts them back to all nodes.
In AR (\Cref{fig:reduce-arch}b), instead of having separate servers, we distribute the aggregation task across workers, each reducing a subset of gradients and distributing them among themselves (\eg, Ring~\cite{patarasuk2009bandwidth}).
Both these architectures have their pros and cons. 
PS operates well in environments with less powerful worker machines but is bandwidth hungry---increasing linearly with the number of worker nodes. 
AR, especially Ring, is bandwidth-optimal but leads to longer execution delays that inflate with the number of worker nodes.

\begin{figure}
     \centering        
     \includegraphics[width=0.95\linewidth]{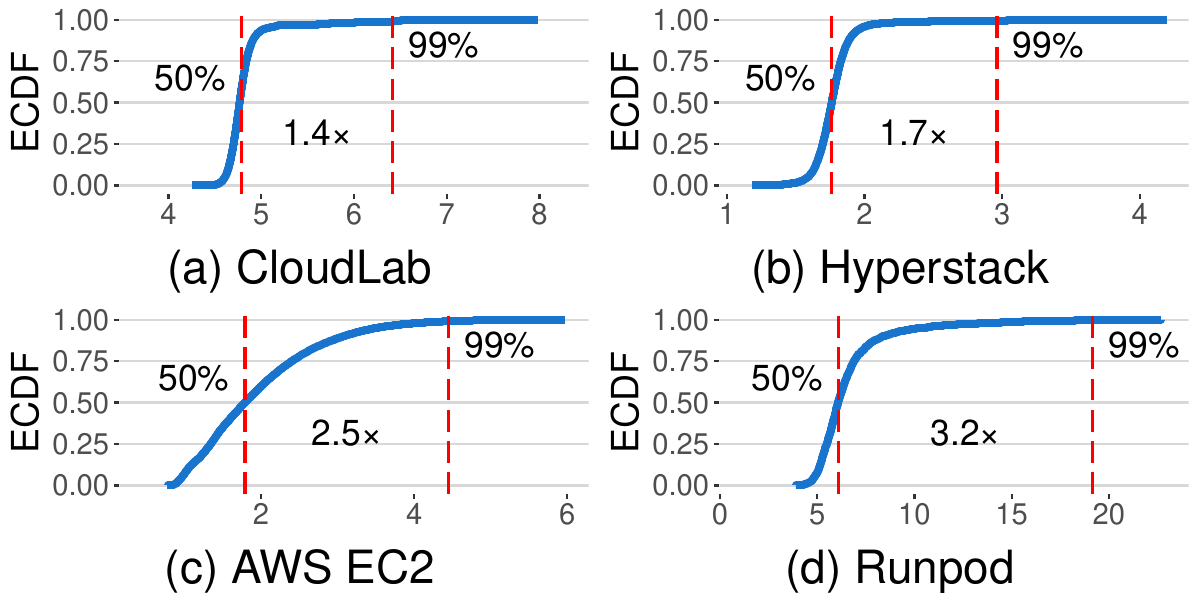}
     \vspace{-8pt}
     \caption{{\bf The latency ECDF (in milliseconds) showing tail-to-median ratio ($\bm{P_{99/50}}$) observed across leading AI cloud platforms.}}
     \vspace{-18pt}
     \label{fig:cloud-tail}
\end{figure}

\vspace{-3pt}
\paragraph{Impact of Stragglers on Performance.}
As shown in \Cref{fig:epochs}, during each backward pass, all DDP worker nodes wait for the gradient aggregation (GA) operations to complete before processing the next batch of data.
The forward and backward passes computation mostly takes place on a machine-learning (ML) accelerator (\eg, GPU or TPU)---a highly parallel and pipelined architecture with predictable and bounded execution time~\cite{williams2009roofline,xu2022igniter}.
Therefore, it is typically the GA operations that lead to long tails and GPU stalls (taking as much as 50\% of the overall DDP processing time)~\cite{sapio2021scaling,shah2023taccl}.
Our measurements across major AI cloud platforms---including AWS EC2~\cite{aws}, Hyperstack~\cite{hyperstack}, CloudLab~\cite{cloudlab}, and RunPod AI~\cite{runpod}---quantify network tail latencies in a distributed training environment. 
Using the Gloo benchmark~\cite{gloo-bench} with 2K gradients on eight nodes, we observe tail-to-median ($P_{99/50}$) latency ratios reaching up to 3.2$\times$ (\Cref{fig:cloud-tail}).

Various factors can contribute to this slowdown in gradient aggregation, including slow workers, transmission delays, incast effects, packet loss and retransmissions, network congestion, and more.
For example, even in the PS architecture, each node sends a complete set of gradients to the parameter server, which can result in excessive drops and retransmissions, due to high incast at the ToR switch~\cite{zhang2017poseidon,li2014communication}---hence, increasing the time to process gradients.
Similarly, in Ring, a single slow worker (or a buggy link) can cause significant delays, because all nodes participate in the aggregation operation in the form of a ring.

\vspace{-2pt}
\subsection{Straggler Mitigation \& Gradient Loss}
\label{ssec:stragglers-loss}
Mitigating stragglers in distributed systems (as well as distributed deep learning) is an active area of research~\cite{ananthanarayanan2013effective,crankshaw2017clipper,dean2013tail,tandon2017gradient,karakus2017straggler,harlap2016addressing,chen2016revisiting,rigazzi2019dc,yakimenka2022straggler,xia2019rethinking}.
One direction focuses on treating stragglers as black boxes and employs schemes, such as redundant task execution~\cite{xiong2021straggler,ghobadiflexent} or skipping slow workers~\cite{cipar2013solving,yakimenka2022straggler,dutta2018slow}, to mitigate the delays due to network congestion or heterogeneous hardware, for example.
They either employ backup workers and select the output of the fastest ones, or simply skip the slow workers altogether.
However, the former can significantly increase the operational expense (in dollars). 
For example, training a GPT-3 model consisting of 175 billion parameters over 355 GPU-years (on a V100)~\cite{awsp3,brown2020language} can cost an additional \$1 Million (\$5.6 Million total) on the AWS instance, \verb|p3.16xlarge|,~\cite{lambda} with only 16\% backup nodes (\ie, 2 backups for every 10 worker nodes).
This cost can further inflate by about 10$\times$ when using more powerful GPUs (\eg, A100 and H100)~\cite{awsp4, awsp5}, higher link speeds (\eg, 40/\SI{100}{Gbps}), and RDMA-enabled NICs~\cite{awsefa}.
Whereas ignoring worker nodes entirely, in the latter case, can lead to slower convergence rates and poor accuracies~\cite{cipar2013solving,yakimenka2022straggler}.
The other direction is to replace commodity servers with specialized hardware (\eg, powerful machines with predictable performance~\cite{zhang2014fuxi,yadwadkar2012proactive}) and dedicated (lossless) communication fabric~\cite{alizadeh2010data,ko2008case,pfister2001introduction}.
Despite their success in HPC-like environments~\cite{petrini2003case,murray2013naiad,xu2013bobtail,ouyang2019mitigating}, these solutions are not applicable in a cloud environment with myriads of tenants, all sharing the resources of the underlying data centers. 

Instead of treating them as black boxes or specialized devices, we argue to replace the (tail-prone) deterministic, run-to-completion workers with their best-effort, \emph{time-bounded} implementations.
The idea is to restrict the processing time of a slow worker and utilize its partial output (gradients) in the next training phase, rather than skipping it entirely. 

\begin{figure}
  \centering  
  \includegraphics[width=0.88\linewidth]{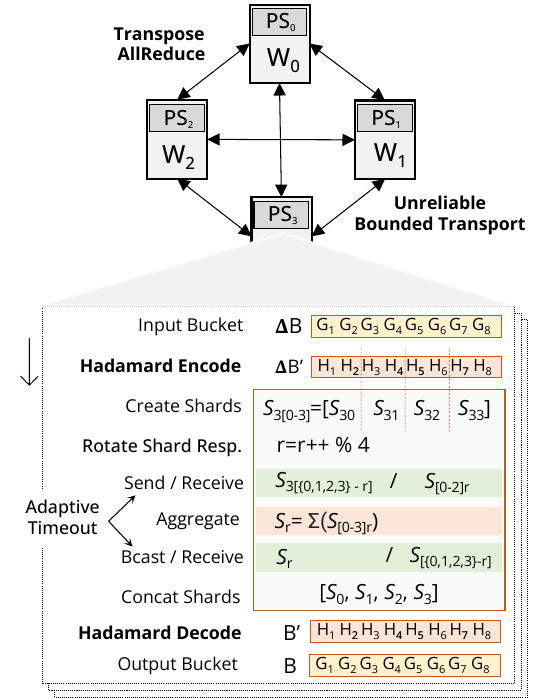}
  \vspace{-4pt}
  \caption{\bf The \name{} design: Transpose AllReduce with colocated parameter servers, Unreliable Bounded Transport, and Hadamard Transform.}
  \label{fig:ultima-design}
  \vspace{-18pt}
\end{figure}

\vspace{-3pt}
\paragraph{Resilience to Gradient Loss vs. Performance and Accuracy.}
Unlike traditional distributed systems (\eg, file sharing and web serving), the stochastic nature of distributed deep-learning systems provides an interesting trade-off between gradient loss (approximation or drops), performance, and accuracy.
These systems (based on SGD-based optimization)~\cite{bottou2012stochastic} are shown to be resilient against estimation inaccuracies in stochastic gradients under different settings~\cite{gupta2015deep, wang2024towards}.
For example, various gradient sparsification~\cite{wangni2017gradient,fei2021efficient,renggli2019sparcml} and quantization~\cite{lin2017deep,alistarh2017qsgd,wen2017terngrad} schemes employ this fact to reduce network traffic overhead.
ATP~\cite{lao2021atp} and SwitchML~\cite{sapio2021scaling} utilize fixed-point arithmetic to execute gradient aggregation in programmable switches with acceptable approximation loss.
Hardware designers incorporate approximate operations (\eg, approx. multipliers~\cite{10.1145/3394885.3431632, rucker2021chopping}) to minimize chip area and energy usage.
More recently, MLT~\cite{wang2024towards} demonstrated that these deep-learning models (\eg, CNNs and LMs) are also resilient to a certain degree of gradient drops---sustaining high accuracy up to 1\% of gradient loss.
Additionally, the impact of gradient loss further diminishes as the number of worker nodes increases~\cite{yu2019distributed}.

\vspace{-2pt}
\section{Design of \name{}}
\label{sec:design}

We present \name{}, a robust AllReduce communication collective system optimized to mitigate tail-latency by exploiting the unique characteristics of distributed-deep learning (DDL)---\ie, resiliency to gradient loss---to quickly reach the convergence accuracies of traditional architectures (\eg, PS and Ring) while mitigating the impact of stragglers and network variabilities.

\vspace{-3pt}
\paragraph{Overview.} 
\Cref{fig:ultima-design} shows the various components of \name{}. 
{\em Transpose AllReduce (TAR)} (\S\ref{sec:design:tar}) implements a peer-to-peer collective-communication fabric, where each node also serves as a parameter server (PS)---a colocated PS architecture~\cite{jiang2020unified, fei2021efficient}. 
{\em Unreliable Bounded Transport (UBT)} (\S\ref{sec:design:bt}) allows these nodes to connect with each other in a best-effort but controlled manner, and bounds the time spent by the two send(bcast)/receive stages during the AllReduce phase.
The PS nodes encode (and decode) gradients in the input bucket ($B$), using {\em Hadamard Transform (HT)} (\S\ref{sec:design:ht}) to disperse the effect of gradient loss, before creating shards ($S_{ij}$) of gradients to be sent to other nodes for reduction. 
They iteratively {\em rotate shard responsibility} among themselves by maintaining a global index ($r$).
Finally, \name{} employs mechanisms (\ie, snapshots and selective skipping) to protect against excessive gradient loss due to transient errors or failures (\S\ref{sec:design:safeguard}); these safeguards ensure robustness and help maintain accuracy under unstable conditions.
All these components operate in tandem to optimize for three competing objectives in \name{}: maximizing performance, minimizing gradient drops, and sustaining accuracy.

\begin{figure*}[t]
  \centering
  \includegraphics[width=0.95\linewidth]{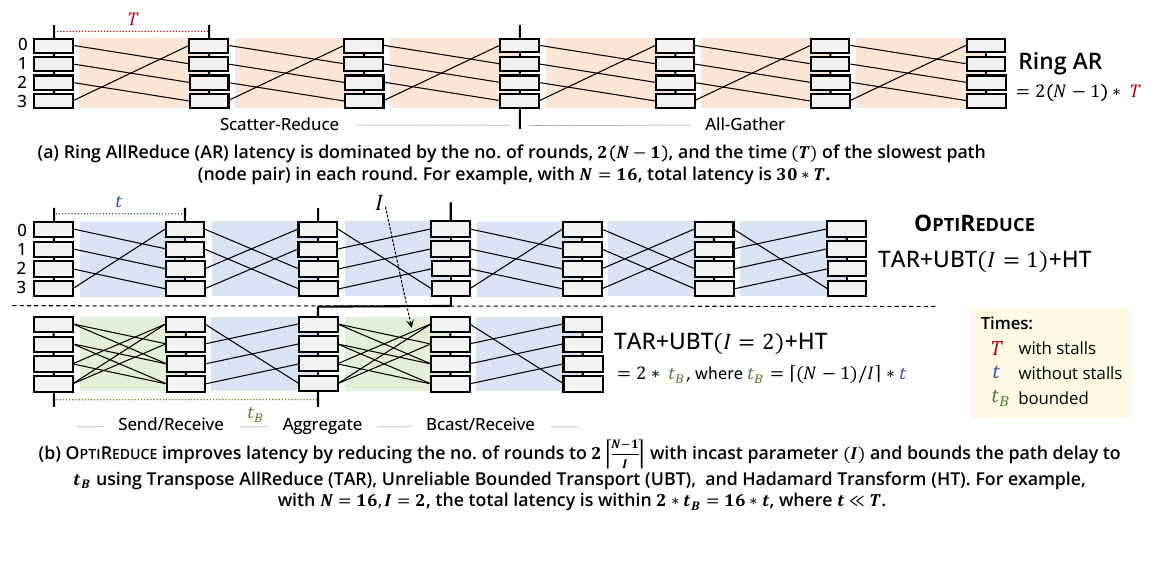}
  \vspace{-32pt}
  \caption{\bf A comparison of Ring versus \name{}.}
  \vspace{-18pt}
  \label{fig:ultima-insights}
\end{figure*}

\vspace{-2pt}
\subsection{Transpose AllReduce (TAR)} 
\label{sec:design:tar}

We begin with Transpose AllReduce (TAR), which implements a hierarchical peer-to-peer gradient-sharing strategy to limit the impact of lost gradient entries when applying the other design optimizations, discussed later.
TAR operates by having each node send its gradient entries directly to all other nodes during an AllReduce phase for aggregation; hence, a lost entry would only impact the aggregated results of a given node-pair in that phase.
Whereas, in Ring~\cite{patarasuk2009bandwidth}, the impact is accumulated and propagated through a ring until it reaches the intended destination nodes.
For instance, in our microbenchmarks (\S\ref{sec:micro}), the Mean Squared Error (MSE) between the expected gradients and those of Ring in the presence of loss is 6$\times$ that of TAR.

\vspace{-2pt}
\subsubsection{TAR Algorithm: A Colocated PS-inspired Collective.~}
\label{tar:design}

TAR combines the key features of traditional AllReduce (\ie, P2P communication)~\cite{clarke1994mpi,li2015malt} and Ring (\ie, minimizing bandwidth using shards and avoiding incast via rounds)~\cite{patarasuk2009bandwidth}.

\begin{figure}[t]
  \centering
  \includegraphics[width=0.92\linewidth]{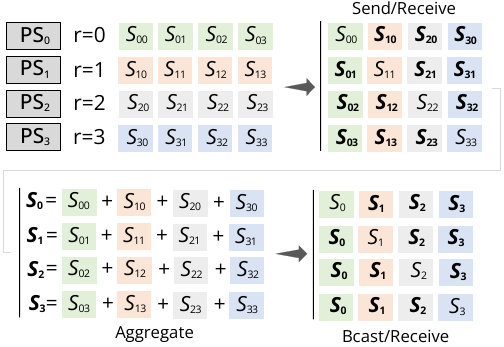}
  \vspace{-8pt}
  \caption{\bf Transpose AllReduce algorithm: PS nodes send/receive shards $S_{ij}$, aggregate them, and bcast/receive to other nodes (all acquiring the same copy).}
  \vspace{-18pt}
  \label{fig:transpose-algo}
\end{figure}

In TAR, each node acts as a worker as well as a parameter server (PS), connected together over a P2P collective-communication fabric (\Cref{fig:ultima-design}).
The $i^{th}$ PS node ($PS_i$) receives a bucket of gradient entries ($\Delta B$) as input from the worker process ($W_i$) and divides it into $N$ shards ($S_{i[0 \ldots N-1]}$), equal to the number of nodes.
Keeping the $r^{th}$ shard it is responsible for aggregating ($S_{ir}$), the node $PS_i$ sends the remaining shards ($S_{i[\{0 \ldots N-1\}-r]}$) to the neighboring nodes. 
At the same time, $PS_{i}$ waits for its shards ($S_{[\{0 \ldots N-1\}-i]r}$) and aggregates (\ie, averages) them with $S_{ir}$ into a single shard $S_r$.
Next, $PS_{i}$ broadcasts $S_r$ to all nodes and receives the aggregated shards from them, $S_{[\{0...N-1\}-r]}$.
Finally, these shards are concatenated into a bucket ($B$) and forwarded to the worker ($W_i$) to process the next batch of data.
When $r=i$, the whole operation appears like a row-wise sum of the transpose of the shard matrix $S$, as shown in \Cref{fig:transpose-algo}; hence, the name {\em Transpose} AllReduce.

In P2P, each PS node communicates directly with the other nodes (instead of forming a ring). 
When sharing gradients, they all interact with each other twice: once sending/receiving shards to the other node and then broadcasting/receiving aggregated shards, hence limiting the impact of accumulated gradient loss.
Sharding also alleviates the load on the PS nodes, where each node only aggregates a bucket ($B$) worth of gradients (rather than $N*B$).\footnote{PyTorch and TesnorFlow typically use a bucket size of \SI{25}{MB}~\cite{li2020pytorch,romero2022accelerating}.} 
Moreover, TAR utilizes the same bandwidth as Ring by sending $B*(N-1)$ bytes over the network during the two send(bcast)/receive stages.
Lastly, to mitigate incast, TAR splits the communication between PS nodes over multiple rounds, where---unlike Ring with fixed node-pairs (\Cref{fig:ultima-insights}a)---nodes communicate with each other using a round-robin strategy, ensuring a given node-pair never repeats across rounds (\Cref{fig:ultima-insights}b).

\vspace{-2pt}
\subsubsection{Hierarchical 2D TAR: Scaling to Larger Node Clusters.~}
\label{2dtar:design}

TAR scales efficiently using a hierarchical design similar to the 2D Ring~\cite{tree, ueno2019exhaustive}.
By grouping nodes, it first performs intra-group communication, in parallel, to locally aggregate gradients, followed by inter-group communication for global aggregation.
Doing so reduces the number of rounds from $2(N-1)$ in traditional TAR to $2(N/G-1)+(G-1)$, where $N$ is the total number of nodes and $G$ is the number of groups. 
For example, with $N = 64$ and $G = 16$, traditional TAR requires an order of magnitude more rounds than 2D TAR---126 compared to just 21.
We provide more details in~\Cref{appendix:hiertar}.

\vspace{-3pt}
\paragraph{Summary:} TAR functions similarly to Ring, yet it {\em reduces the impact of lost gradients} by establishing P2P communication among all nodes in each round, avoiding the propagation of losses via aggregation through intermediate nodes.

\vspace{-2pt}
\subsection{Unreliable Bounded Transport (UBT)}
\label{sec:design:bt}

One of the primary causes affecting tail latency in DDL is the variability in the network behavior due to congestion~\cite{dean2013tail,wang2024towards}.
Current transport protocols (like TCP) further exacerbate these effects by demanding reliable, in-order delivery of packets (gradients) between training nodes---if packets are dropped or received out-of-order, TCP will stall until all gradients are received over the affected path.

\begin{figure}[t]
\centering
\includegraphics[width=0.92\linewidth]{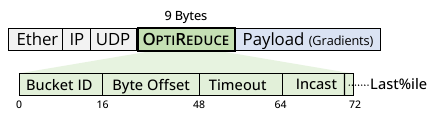}
\vspace{-12pt}
\caption{\bf \name{}'s header format}
\label{fig:ultima-header}
\vspace{-18pt}
\end{figure}

However, simply replacing TCP with message-based protocols (like UDP) would not work.
While UDP is faster, as it avoids packet retransmission and reordering, it lacks congestion control, which can lead to network congestion collapse.
Moreover, UDP sends data at full link speed (\eg, \SI{100}{Gbps}), causing excessive drops (loss of gradients) beyond what DDL models can tolerate.

To address this, we enhance UDP with adaptive timeouts, dynamic incast, and minimal rate control to create a new Unreliable Bounded Transport (UBT) protocol, which limits computation and communication time while maximizing gradient delivery in each round.
It adds a new 9-byte header, \verb|OptiReduce| (\Cref{fig:ultima-header}), to commit arriving packets (with gradients) to the right bucket and offset using the header fields, \verb|Bucket ID| and \verb|Byte Offset|, respectively.
These fields ensure that gradients reach the correct bucket, irrespective of the ordering of the incoming packets when multiple GA operations are running in parallel (\Cref{fig:epochs}).

\vspace{-2pt}
\subsubsection{Adaptive Timeout.~}
UBT implements adaptive timeouts to bound the tail communication time of the send(bcast)/receive stages of the GA operations to $t_B$ (\Cref{fig:ultima-insights}b).
By restricting the time to $t_B$, we can control the worst-case execution of these stages---allowing GA operations to finish within a bounded time.

However, there are a couple of challenges with this approach. 
(1) How to select the value of $t_B$? 
Too small will lead to undue loss, and too high will cause unnecessary delays. 
Moreover, the value will vary with environmental settings (\eg, GPU type, CPU clock, vCPUs, and interface speed) and parameters (\eg, no. of nodes, bucket sizes, and incast).
(2) A single lost packet, which is likely in UBT, would cause the GA operation to always take $t_B$ (worst-case) time to finish.
TCP, on the other hand, can perform better in some instances where communication may finish faster than waiting for the full timeout ($t_B$), even with retransmissions.

\vspace{-3pt}
\paragraph{Selecting the Timeout Value ($t_B$).}
As shown in \Cref{fig:epochs}, during backpropagation, multiple GA operations execute in parallel on buckets of varying sizes.
For selecting $t_B$, during the initialization phase, we run GA with TAR and TCP, using the largest bucket, for a couple of iterations to collect completion times for both send(bcast)/receive stages.
PS nodes share these values with each other using the \verb|Timeout| field in the \name{} header (\Cref{fig:ultima-header}).

We then form a list of these times and set $t_B$ to the 95th \%ile of that list.
In \S\ref{sec:evaluation}, we show that using 20 iterations and the 95th \%ile value allows \name{} to sustain full model accuracies while finishing up to 2$\times$ faster.

\begin{figure}
\centering
\includegraphics[width=\linewidth]{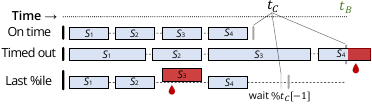}
\vspace{-20pt}
\caption{\bf Different timeout strategies in \name{}.}
\label{fig:adaptive-timeouts}
\vspace{-18pt}
\end{figure}

\vspace{-3pt}
\paragraph{Progressing Quickly via Early Timeout.}
To avoid approaching $t_B$ every time a loss happens, we introduce an early timeout scheme, which causes GA's receive stages to expire whenever there are no remaining gradient entries to read (\ie, the buffer is empty). 
For each bucket, we track a moving average ($t_C$) of completion times; we keep separate averages for both the receive stages in GA (\Cref{fig:ultima-insights}).
The sender PS node tags the last 99th \%ile packets by setting the \verb|Last%ile| field in the header.
When the buffer is empty, the receiver node checks if some of the last \%ile packets have been received from all nodes. 
If so, it waits for an $x$\% of $t_C$ time before expiring (\Cref{fig:adaptive-timeouts}).

The value of $x$\% is dynamically adjusted based on the percentage of gradient entries dropped from the previous round.
Starting at 10\%, the goal is to maintain gradient losses between 0.01\% and 0.1\%. 
If losses exceed this range, $x$\% is doubled until they return within the limit. 
If losses drop below 0.01\%, $x$\% is decreased by 1 until the desired range is reached.
(The maximum $x$\% is capped at 50\%.)
If gradient losses exceed 2\% at any point, we activate Hadamard Transform (\S\ref{sec:design:ht}) to mitigate the effects of dropped gradients on convergence accuracy.
\footnote{The 2\% threshold is set based on prior work~\cite{wang2024towards} and our evaluations~\S\ref{sec:micro}.}

We calculate $t_C$ in the following steps. First, we compute the (expected) completion time of a given receive stage: (1) if on time, then we set $t_C$ to the current time spent, (2) if timed out, then $t_C$ = $t_B$, and (3) if last \%ile received, then $t_C$ is set to the expected time needed to receive all data (\eg, $t_C$ = current time spent $\times$ total / received data).
Next, we pick the median $t_C$ from the values computed by the $N$ PS nodes (shared over the \verb|Timeout| field in the header).
Finally, we calculate the moving average: $t_C$ = $\alpha*t_C + (1-\alpha)*t_C[-1]$.

\vspace{-2pt}
\subsubsection{Dynamic Incast.~}
UBT further introduces a notion of dynamic incast (\Cref{fig:ultima-insights}b). 
The TAR's P2P communication model lets \name{} alter the number of senders ($I$) a PS node can receive gradients from in a given round.
For example, setting $I=1$ (a single sender) would cause TAR to take the same number of rounds as Ring, $2(N-1)$; however, increasing $I=2$ would quickly reduce these rounds by about half, $2\lceil(N-1)/2\rceil$; and so on.

The incast parameter ($I$) can be configured either statically at boot time, based on the available network and node capacity (\eg, modern datacenters can handle hundreds of thousands of incast packets without degrading performance~\cite{handley2017re,montazeri2018homa}), or dynamically adjusted based on runtime metrics (like throughput, latency, or loss rate). 
In UBT, receivers dynamically modify the incast factor in response to current loss and timeout events. If the loss rate increases, the factor is reduced to alleviate congestion; conversely, if the loss rate remains low (indicating timely packet reception with no timeouts), the incast factor is increased. Receivers communicate their incast factor, $I$, by updating the \texttt{Incast} field in the \name{} header (\Cref{fig:ultima-header}), and the sender then selects the smallest reported value of $I$ for that round.

\begin{figure}
 \centering
 \includegraphics[width=\linewidth]{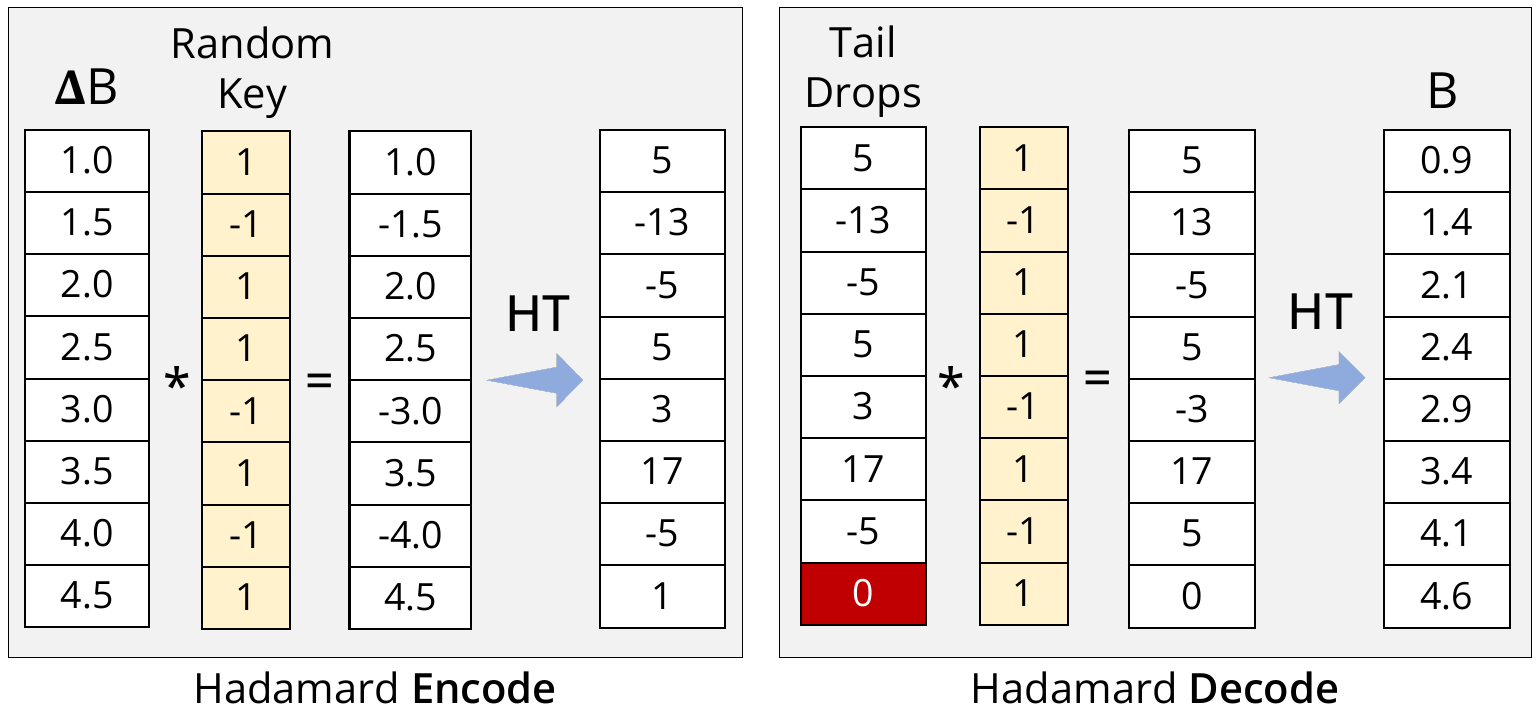}
 \vspace{-18pt}
 \caption{\bf Dispersing the effect of lost gradients (\eg, due to tail drops) using Hadamard Transform (HT).}
 \label{fig:hadamard}
 \vspace{-18pt}
\end{figure}

\vspace{-2pt}
\subsubsection{Minimal Rate Control.~}
Since \name{} is resilient to loss, we only require a minimal scheme for rate control to prevent congestion collapse.
For that, UBT employs a basic TIMELY-like rate-control mechanism~\cite{mittal2015timely}, where the sender adjusts flow rates based on RTT feedback derived from timestamps returned by the receiver at regular intervals (every 10th packet) over a separate control channel. 
If the RTT (or its gradient) is below $T_{low}$, the sender increases the rate by $\alpha$, and if the RTT exceeds $T_{high}$, the rate is reduced by $(1-\beta \cdot (1-T_{high}/\text{RTT}))$.
In our experiments, we set $T_{low} = 25\mu$s, $T_{high} = 250\mu$s, $\alpha = 50$ Mbps, and $\beta = 0.5$, when running in a shared environment~\cite{mittal2015timely,wang2024towards}.

\vspace{-3pt}
\paragraph{Summary:} UBT, in conjunction with TAR, {\em improves tail latency} by minimizing the impact of network congestion; adaptive timeouts bound the latency of the send(bcast)/receive stages, while dynamic incast reduces the number of communication rounds.

\subsection{Dispersing Gradient Loss} 
\label{sec:design:ht}

Finally, to make \name{} resilient against drop patterns (\eg, tail drop) in the network, we employ {\em randomized} Hadamard Transform (HT)~\cite{vargaftik2022eden, vargaftik2021drive, suresh2017distributed}, which spreads the effect of a dropped gradient over the entire bucket.
For example, in \Cref{fig:hadamard}, HT encodes a bucket ($\Delta B$) and sends it over the network.
Upon reception, the last gradient (in red) was lost; however, HT preserves the lost information by slightly perturbing the values of other gradients in the decoded bucket $(B)$.
The Mean Squared Error (MSE) between the decoded and received (without HT) bucket, compared to the original one, is 0.01 and 2.53, respectively.
That is why, when combined with rotating shard responsibility between nodes (\Cref{fig:ultima-design}), HT lets \name{} be more aggressive with the timeout value ($t_B$) while still reaching high model convergence accuracies (\S\ref{sec:evaluation}).

\vspace{-3pt}
\paragraph{Summary:}
HT, together with TAR+UBT, {\em limits the effect of dropped gradients} by spreading it across the entire bucket, thus preserving the lost information. 
Additionally, it allows \name{} to operate faster, with stringent $t_B$ values, without affecting convergence accuracies (\S\ref{sec:evaluation}). 

\vspace{-2pt}
\subsection{Safeguards against Excessive Loss}
\label{sec:design:safeguard}
\name{} continuously monitors gradient loss during each AllReduce phase, and if the loss exceeds a predefined threshold, it can either skip the gradient update for that round or automatically halt the training, prompting user intervention.
Skipping an update helps minimize potential harm to the overall training process by discarding transient high-loss updates without impacting long-term model accuracy or completion time. 
This mechanism helps prevent major disruptions in the training process, ensuring users are notified of any accuracy concerns and can make necessary adjustments.
Similar techniques are routinely integrated into modern deep-learning pipelines to monitor, track, and recover model accuracy~\cite{nigenda2022amazon,evidentlyai,neptuneai,torchft}.

\vspace{-2pt}
\section{Implementation}
\label{sec:implementation}

We develop \name{} as a new collective-communication scheme inside the Gloo library (v0.5.0)~\cite{gloo} and integrate it with PyTorch Distributed (v1.12)~\cite{li2020pytorch}, a widely used deep-learning framework---allowing \name{} to work without modification with a large body of deep-learning models (\eg, CNNs~\cite{krizhevsky2017imagenet, simonyan2014very, he2016deep}, RNNs~\cite{hochreiter1997long,mikolov2010recurrent,cho2014learning}, and Transformers~\cite{vaswani2017attention,devlin2018bert,brown2020language}).
We pick Gloo due to its simpler design and our familiarity with the codebase; however, we expect \name{} will yield similar benefits when operating with other popular libraries (\eg, NCCL~\cite{jeaugey2017nccl} or MSCCL~\cite{msccl}).

We extend the C++ implementation of Gloo to support our Transpose AllReduce (TAR) collective and provide support for both reliable transport (over TCP) and our best-effort transport (over UBT).
We prototype UBT as a userspace transport layered on UDP, including rate control, using Nvidia DPDK API (v20.11)~\cite{dpdk}.

We further add support for communication hiding in \name{}, \ie, running two AllReduce operations in parallel with backpropagation.\footnote{This is consistent with existing parallelism approaches that allow for two concurrent AllReduce operations (\eg, PyTorch~\cite{li2020pytorch}).}
The sender maintains separate layer-3 port numbers to tag gradients for the two parallel AllReduce operations. 
On the receive side, two PMD threads poll incoming traffic (gradients) in their local receive queues. 
An Nvidia Connectx-6 NIC routes traffic to the respective queues based on the port numbers; we install rules in the NIC using DPDK's \verb|rte_flow| API~\cite{flowrule}.
We also install rules to route non-\name{} traffic to the kernel using DPDK's Flow Bifurcation mechanism~\cite{dpdkbif}.\footnote{Flow Bifurcation is a mechanism that lets hardware-capable NICs forward traffic directly to the userspace (DPDK thread) or the Linux kernel.}
Doing so ensures that Gloo's kernel stack remains unaffected and other network operations, \eg, rendezvous in PyTorch DDP~\cite{li2020pytorch}, continue uninterrupted.

To include support for adaptive timeouts, we use C++ STL library's 
\verb|wait_for()| function~\cite{josuttis2012c++}, which is a blocking call that returns either when a given condition is met (such as received all gradients) or a timeout occurs. 
For the timeout, we pair the \verb|wait_for()| function with Chrono library's \verb|high_resolution_clock()|~\cite{josuttis2012c++} to operate at nanosecond clock granularity.
For Hadamard Transform, we apply a widely-used C++/CUDA implementation by HazyResearch~\cite{structured-nets}, which uses GPUs to perform this operation.
We use PyTorch DDP's communication hook~\cite{ddphook} to register Hadamard's encode/decode callbacks for processing gradient buckets before and after reduction, respectively.
\vspace{-2pt}
\section{Evaluation}
\label{sec:evaluation}

\begin{figure}
  \centering
  \includegraphics[width=\linewidth]{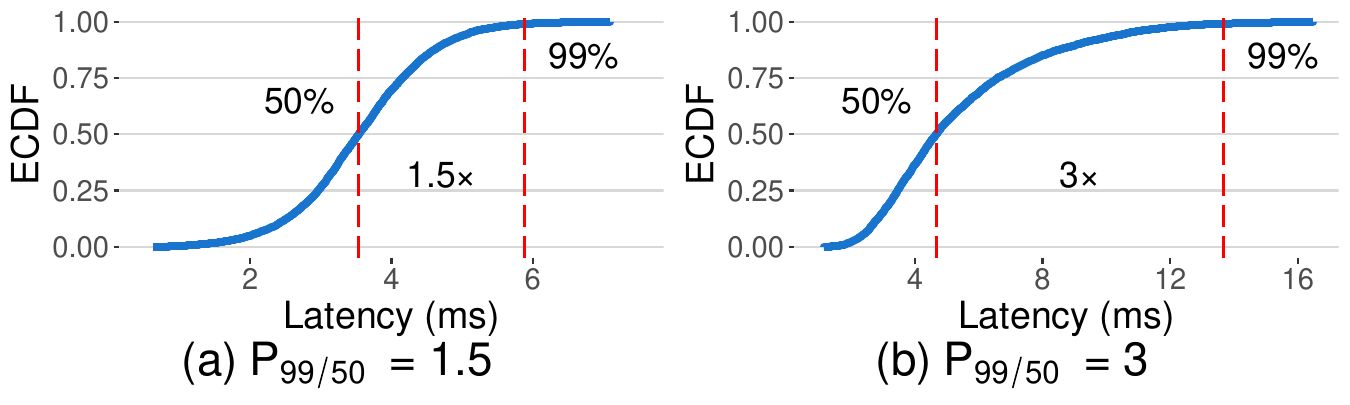}
  \vspace{-20pt}
  \caption{\bf Our local cluster with tail-to-median latency ratio ($\bm{P_{99/50}}$) of 1.5 (a) and 3 (b).}
  \label{fig:latency}
  \vspace{-15pt}
\end{figure}

In this section, we provide an end-to-end comparison of \name{} with state-of-the-art solutions (\S\ref{sec:end2end}), and microbenchmark the utility of its design components (\S\ref{sec:micro}).

\vspace{-2pt}
\subsection{Experimental Setup}
\label{ssec:exp-setup}

\begin{figure*}[t]
    \centering
    \includegraphics[width=0.7\linewidth]{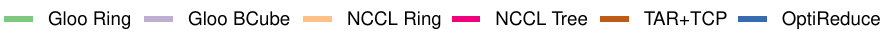}

    \begin{subfigure}[b]{\linewidth}
		\vspace{-2pt}
        \centering        
        \begin{subfigure}[t]{0.32\linewidth}
            \centering
            \includegraphics[width=\textwidth]{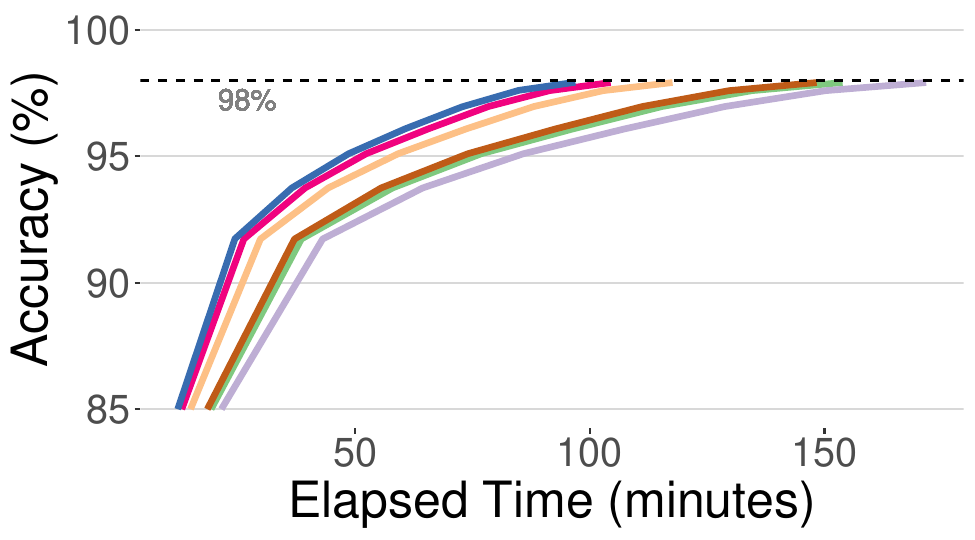}
            \vspace{-18pt}
            \caption{Local Cluster: $\bm{P_{99/50} = 1.5}$}
            \label{fig:cloudlab-gpt2-v100-0-1}
        \end{subfigure}
        \hfill
        \begin{subfigure}[t]{0.32\linewidth}
            \centering
            \includegraphics[width=\textwidth]{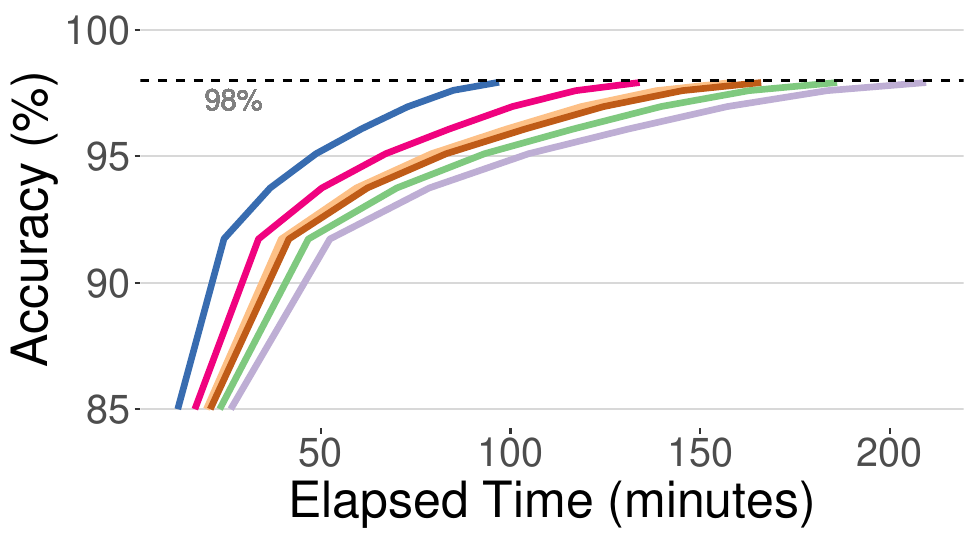}
            \vspace{-18pt}
            \caption{Local Cluster: $\bm{P_{99/50} = 3}$}
            \label{fig:cloudlab-gpt2-v100-1}
        \end{subfigure}
        \hfill
        \begin{subfigure}[t]{0.32\linewidth}
            \centering
            \includegraphics[width=\textwidth]{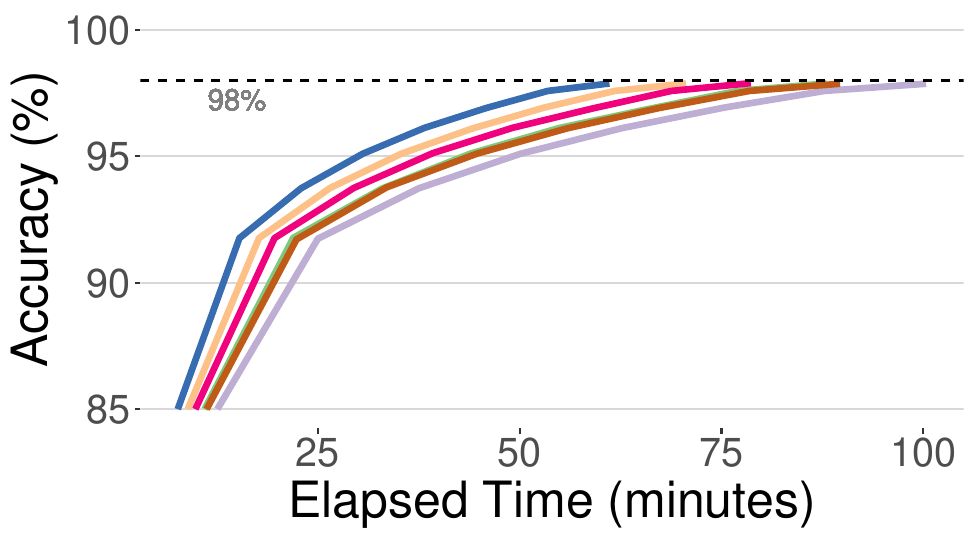}
            \vspace{-18pt}
            \caption{CloudLab}
            \label{fig:cloudlab-gpt-a30}
        \end{subfigure}
        
    \end{subfigure}
    \vspace{-8pt}
    \caption{\bf Time-to-accuracy (TTA) comparison for the OpenAI GPT-2 model with eight worker nodes.}
    \vspace{-8pt}
    \label{fig:gpt2-tta}
\end{figure*}

\begin{figure*}[t]
    \centering
    \includegraphics[width=0.6\linewidth]{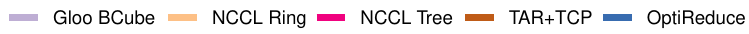}
	
    \begin{subfigure}[b]{\linewidth}
		\vspace{-2pt}
        \centering
        \begin{subfigure}[t]{0.32\linewidth}
            \centering
            \includegraphics[width=\textwidth]{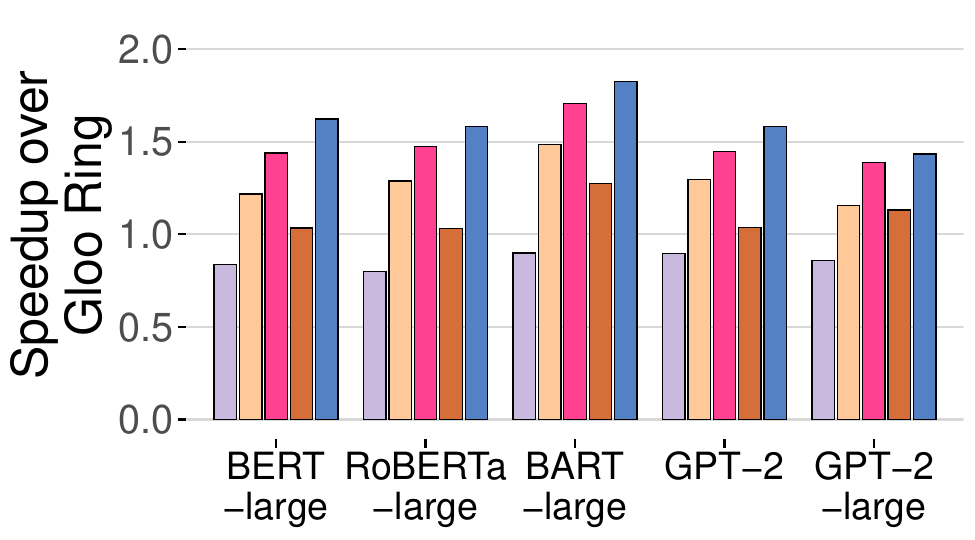}
            \vspace{-18pt}
            \caption{Local Cluster: $\bm{P_{99/50} = 1.5}$}
            \label{fig:cloudlab-tta-v100-0-1}
        \end{subfigure}
        \hfill
        \begin{subfigure}[t]{0.32\linewidth}
            \centering
            \includegraphics[width=\textwidth]{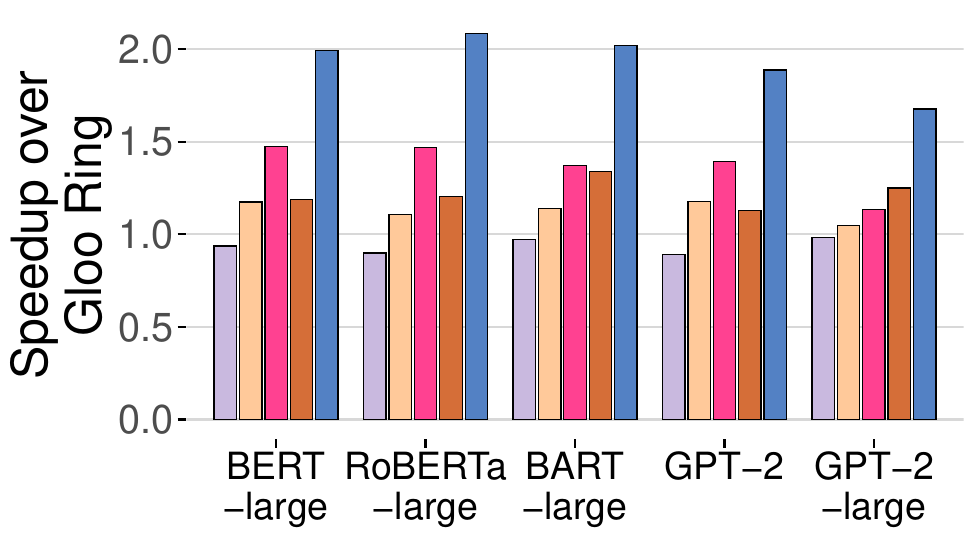}
            \vspace{-18pt}
            \caption{Local Cluster: $\bm{P_{99/50} = 3}$}
            \label{fig:cloudlab-tta-v100-1}
        \end{subfigure}
        \hfill
        \begin{subfigure}[t]{0.32\linewidth}
            \centering
            \includegraphics[width=\textwidth]{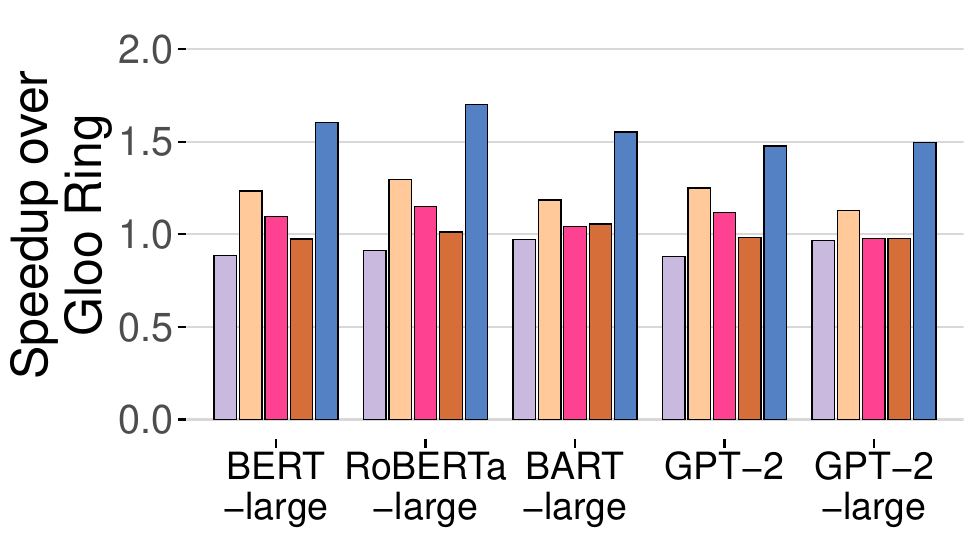}
            \vspace{-18pt}
            \caption{CloudLab}
            \label{fig:cloudlab-tta-a30}
        \end{subfigure}
    \end{subfigure}
    \vspace{-8pt}
    \caption{\bf Training throughput comparison for large language models (LLMs) with eight worker nodes.}
    \vspace{-15pt}
    \label{fig:all-throughput}
\end{figure*}

\subsubsection{Test Environments.~}
We evaluate \name{} in both our local virtualized cluster and a real-world environment using CloudLab~\cite{duplyakin2019design}.

\vspace{-3pt}
\paragraph{a) Local Virtualized Cluster.} 
Our local testbed is a collection of four servers configured as a virtualized cluster~\cite{proxmox}.
Each machine has a 32-core AMD EPYC 7542 CPU @ \SI{2.90}{GHz}, \SI{512}{GB} RAM, two Nvidia Tesla V100 GPUs, and a ConnectX-5 dual-port NIC.
In total, there are eight V100 GPUs, one per VM in the cluster.
The VMs communicate over the network using a dedicated NIC port with Nvidia's OFED device drivers (v24.04).
Both GPU and NIC interfaces are exposed to the VMs via Intel's VT-d PCIe passthrough technology~\cite{abramson2006intel}---allowing direct (dedicated) access to the physical functions.
A programmable switch (Tofino1~\cite{tofino}) connects the servers and VMs over a \SI{25}{Gbps} network. Additionally, it facilitates in-network aggregation for SwitchML benchmarks (\S\ref{sec:micro}).

Recent studies from Microsoft~\cite{shah2023taccl}, Amazon~\cite{karakus2021amazon}, and Google~\cite{dean2012large, dean2013tail} show that the tail-to-median ratio ($P_{99/50}$) for distributed workloads, including deep-learning training, ranges from 1.5$\times$ to 4$\times$ in large cloud data centers~\cite{li2020pytorch, dean2013tail, abbasi2022dwtcp}.\footnote{Even CloudLab, a relatively small-scale cloud compared to commercial ones, exhibits a $P_{99/50}$ ratio of around 1.45.}
To emulate these environments and their tail characteristics in our testbed, we follow the approach of previous studies~\cite{alizadeh2013data,alizadeh2014conga,yan2021acc,addanki2022powertcp,vishwanath2008evaluating} by running background workloads on random nodes and links. 
Varying the number of concurrent workloads allows us to adjust the tail-to-median latency ratio within the network.
We validate the fidelity of our scheme using the Gloo benchmark utility~\cite{gloo-bench} with 2K gradients.
As shown in \Cref{fig:latency}, our method accurately preserves the expected latency distributions, maintaining the $P_{99/50}=1.5,3$ ratios.

\vspace{-3pt}
\paragraph{b) Public Cloud: CloudLab.}
We configured our real environment on CloudLab~\cite{duplyakin2019design}, a public research cloud widely shared by researchers and academics for computing and distributed systems experimentation. 
We provisioned eight \verb|d7525| instances~\cite{cloudlab}, each equipped with an Nvidia Ampere A30 GPU and a ConnectX-6 DX dual-port NIC, all connected via a \SI{10}{Gbps} network.

\vspace{-2pt}
\subsubsection{Baselines, Workloads, and Parameter Settings.~} 
We evaluate \name{} against the following baseline systems: Gloo (Ring~\cite{patarasuk2009bandwidth} and BCube~\cite{guo2009bcube}), NCCL (Ring~\cite{patarasuk2009bandwidth} and Tree~\cite{tree}) with TCP, as well as a reliable version of our Transpose AllReduce (TAR) with TCP (TAR+TCP).
Additionally, we evaluate BytePS and three popular compression algorithms: Top-K~\cite{stich2018sparsified}, TernGrad~\cite{wen2017terngrad}, and THC~\cite{li2024thc}. 
To provide further insights, we also microbenchmark \name{} against in-network systems such as SwitchML~\cite{sapio2021scaling}, despite their reliance on switch-level access within the provider's network, which makes them inapplicable for cloud environments.

We train a variety of language models (LMs), including BERT-base/large~\cite{devlin2018bert} and RoBERTa-base/large~\cite{liu2019roberta} on the SQuAD 2.0 dataset~\cite{rajpurkar2016squad}, as well as BART-base/large~\cite{lewis2019bart} and OpenAI GPT-2-base/large~\cite{radford2019language} on the GLUE benchmark~\cite{wang2018glue} for the SST2 (Stanford Sentiment Treebank) task~\cite{socher2013recursive}.
We further evaluate \name{} on additional models and tasks, which we discuss in \Cref{appendix:llama} and~\ref{appendix:int-llms}. 
Specifically, we train the Llama-3.2 1B model~\cite{dubey2024llama} on three standard downstream tasks: SQuAD (extractive question answering)~\cite{rajpurkar2016squad}, ARC (science reasoning)~\cite{clark2018think}, and MATH (symbolic mathematics)~\cite{hendrycks2021measuring} (\Cref{appendix:llama}).
Additionally, we evaluate and microbenchmark \name{} on network-intensive models (VGG-16/19)~\cite{simonyan2014very} using the CIFAR-100 dataset~\cite{krizhevsky2009learning} and compute-intensive models (ResNet-50/101/152)~\cite{he2016deep} with the ImageNet dataset~\cite{russakovsky2015imagenet} (\Cref{appendix:int-llms}).

We compute the \name{}'s timeout value ($t_B$) for each model using 20 iterations; we set $\alpha=0.95$ when calculating the moving average ($t_C$).
We use the incast parameter of $I=1$, unless stated otherwise.

\begin{table*}
\begin{center}

    \small
    \begin{tabular}{l|cc|cc|c|>{\columncolor[HTML]{cfe2f3}}c>{\columncolor[HTML]{cfe2f3}}c}
        \toprule
        \multirow{2}{*}{{\bf Test Environment}} 
        & \multicolumn{2}{c|}{{\bf Gloo}} & \multicolumn{2}{c|}{{\bf NCCL}} & \multicolumn{1}{c|}{\multirow{2}{*}{{\bf TAR+TCP}}} & \multicolumn{1}{c}{\cellcolor[HTML]{cfe2f3}{\color[HTML]{333333}}} & \multicolumn{1}{c}{\multirow{1}{*}{\cellcolor[HTML]{cfe2f3}{\color[HTML]{333333}{\em Dropped Gradients}}}} \\ \cline{2-5}
                                         & \multicolumn{1}{c|}{\textbf{\em Ring}} & \multicolumn{1}{c|}{\textbf{\em BCube}} & \multicolumn{1}{c|}{\textbf{\em Ring}} & \multicolumn{1}{c|}{\textbf{\em Tree}} & \multicolumn{1}{c|}{}                 & \multicolumn{1}{c}{\multirow{-2}{*}{\cellcolor[HTML]{cfe2f3}{\color[HTML]{333333}{\bf \name{}:}}}} & \multicolumn{1}{c}{\multirow{1}{*}{\cellcolor[HTML]{cfe2f3}{\color[HTML]{333333}{\em (\%Entries)}}}} \\ \toprule
        \multirow{1}{*}{Local Cluster: $P_{99/50} = 1.5$}              
			& 154 & 172 & 118 & 105 & 148 & 96 & 0.07 \\ 
        \multirow{1}{*}{Local Cluster: $P_{99/50} = 3.0$}                
			& 186 & 210 & 159 & 135 & 166 & 97 & 0.18 \\ 
        \midrule
                \multirow{1}{*}{CloudLab}         
			& 88 & 100 & 71 & 79 & 90 & 60 & 0.05 \\ 
                                         \bottomrule
    \end{tabular}
	\vspace{-5pt}
    \caption{\bf Comparing the end-to-end convergence time (in minutes) of baseline systems vs. \name{} for OpenAI GPT-2 (total gradients, \SI{40}{TB}). TAR+UDP suffers excessive drops, losing up to 30\% of gradients, and fails to converge.}
	\vspace{-20pt}
    \label{tab:eval-overview}
\end{center}
\end{table*}

\vspace{-2pt}
\subsection{End-to-End Evaluation}
\label{sec:end2end}
We conduct end-to-end evaluations in two environments: (1) our local virtualized cluster, with tail-to-median ratios $P_{99/50} = 1.5$ (low variability) and 3 (high variability), and (2) a real public cloud, CloudLab.
We compare \name{} against the baseline systems Gloo (Ring and BCube), NCCL (Ring and Tree), and TAR+TCP; and measure time-to-accuracy (TTA), throughput, gradient drop percentage (in bytes), and the achieved training accuracy.

Our results show that \name{} consistently outperforms the baselines. On our local cluster, we observe time-to-accuracy (TTA) reductions of up to (82\%, 98\%) compared to Gloo (Ring, BCube), and (44\%, 25\%) compared to NCCL (Ring, Tree), respectively. 
These improvements extend to CloudLab, where we see average TTA reductions of up to (47\%, 67\%) over Gloo (Ring, BCube), and (18\%, 32\%) over NCCL (Ring, Tree). 
Furthermore, \name{} achieves the same convergence accuracy as the baselines while limiting gradient entry losses to less than 0.1\% of the total traffic.

\vspace{-3pt}
\paragraph{$\bullet$ TTA and Throughput.}
\Cref{fig:gpt2-tta} illustrates how TTA for the five baselines and \name{} varies under different environments---Local cluster ($P_{99/50}$ = 1.5 and 3) and CloudLab---for the OpenAI GPT-2 model. 
Across all runs, \name{} maintains a lower TTA from the onset.\footnote{We observe that under ideal conditions, with $P_{99/50} = 1$ (no variability), all systems perform similarly (not shown).}
For example, on our local cluster with $P_{99/50}=1.5$ (\Cref{fig:gpt2-tta}a), \name{} converges in 96 minutes, while NCCL Tree takes 105 minutes, and the next best, NCCL Ring, taking 118 minutes.
With $P_{99/50} = 3$, the TTA differences become more pronounced (\Cref{fig:gpt2-tta}b). 
\name{} remains unaffected by the increased variability, maintaining its lead in TTA with a 98\% accuracy and finishing in about 97 minutes. 
In contrast, the baselines experience significant slowdowns, with their TTA inflating by 1.41--2.18$\times$ compared to \name{}. 

We see the same trend on CloudLab (\Cref{fig:gpt2-tta}c), \name{} reaches the convergence accuracy in 60 minutes, whereas it is 71 minutes for NCCL Ring.
Other baselines continue to trail behind \name{}, with NCCL Tree having the next-best TTA of 79 minutes.

We observe similar speedups for \name{} when training other models, including BERT-large, RoBERTa-large, BART-large, and GPT-2-large (\Cref{fig:all-throughput}).

\vspace{-3pt}
\paragraph{$\bullet$ Gradient Drops and Convergence Time.}
We further evaluate the drops in gradient entries and their impact on convergence time (\Cref{tab:eval-overview}). 
In our local cluster with $P_{99/50} = 1.5$, a small percentage of gradient entries is lost (\ie, 0.07\%) due to \name{}'s adaptive timeouts in UBT, causing the system to progress without waiting on stragglers.
These timeouts manifest as dropped gradients in \name{}; whereas baseline systems stall on these stragglers. 
Still, \name{} achieves the same convergence accuracy (98\% for GPT-2) as the baselines but in under 96 minutes, compared to 105 minutes for the next best, NCCL Tree.
When $P_{99/50}$ increases to 3, increased congestion in the network and stragglers cause more gradient entries to be lost, but only slightly (0.18\%), and does not impact \name{}'s training accuracy and convergence time, whereas it inflates NCCL Tree's time to 135 minutes.

Similarly, in CloudLab, \name{} sees a 0.05\% drop in gradient entries, which allows it to reach the convergence accuracy in 60 minutes, compared to its next best, NCCL Ring, taking 18\% longer.

\vspace{-2pt}
\subsection{Microbenchmarks}
\label{sec:micro} 
We now evaluate the effectiveness of the individual design components in \name{}.
We run the VGG-19 model on the CIFAR-100 dataset for these measurements using our local cluster.

\vspace{-3pt}
\paragraph{$\bullet$ \name{}'s TAR topology leads to minimum dropped gradients when using a best-effort transport.}
We compare the number of gradients lost across different AllReduce topologies using our Unreliable Bounded Transport (UBT). 
We measure Mean Squared Error (MSE) to gauge the difference between the original gradients and those received over these topologies for three different schemes on our local cluster with $P_{99/50} = 1.5$: Ring topology in Ring-AllReduce, P2P in PS, and P2P with rounds in TAR, using a \SI{500}{M} tensor.
Ring-AllReduce has the worst MSE (14.55)---an order of magnitude greater than TAR (2.47).
The presence of fixed node pairs in Ring-AllReduce (\S\ref{sec:design:tar}) propagates losses, resulting in a higher deviation from the original gradients.
PS also has a high MSE (9.92) due to excessive incast when all nodes send gradients to the parameter server (PS) simultaneously.
In contrast, TAR avoids this by distributing P2P communication over multiple rounds.

\begin{figure}[t]
 \centering  
 \includegraphics[width=0.9\linewidth]{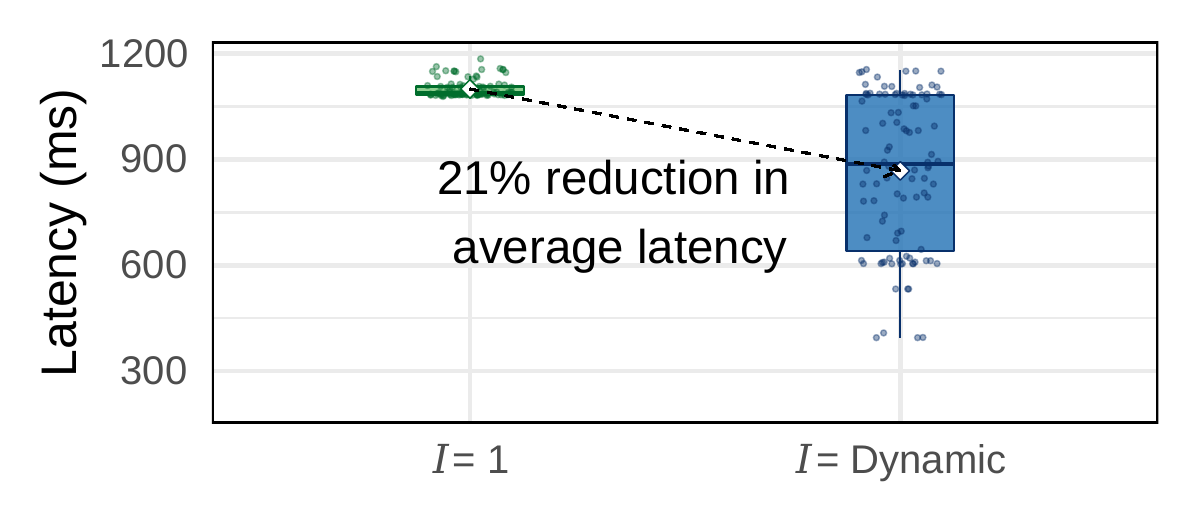}
 \vspace{-15pt}
 \caption{\bf Latency distribution of \name{} with static ($I=1$) vs. dynamic incast feature in UBT, using a synthetic \SI{500}{M}-gradient AllReduce workload.}
 \label{fig:incast}
 \vspace{-15pt}
\end{figure}

\begin{figure*}[t]
     \centering
     \begin{subfigure}[b]{0.32\textwidth}
         \centering         \includegraphics[width=0.95\textwidth]{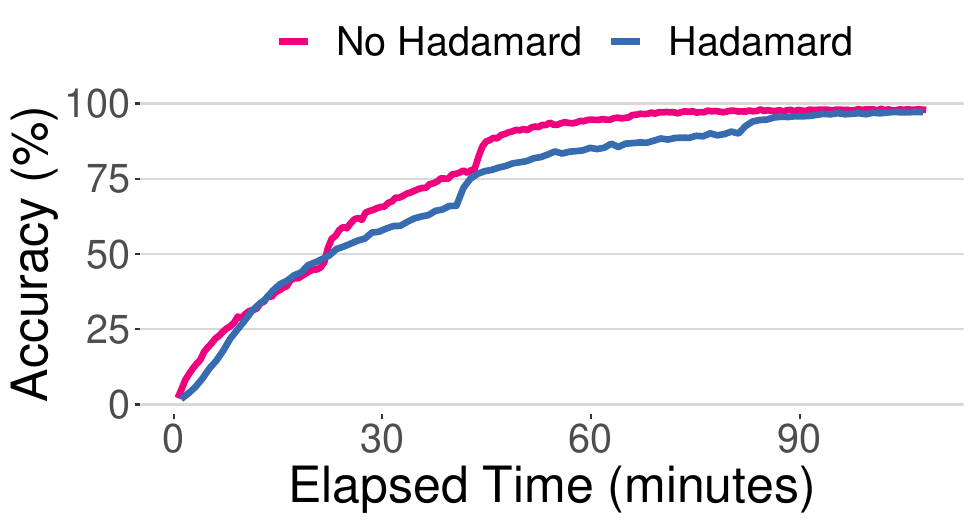}
         \vspace{-5pt}
         \caption{1\% gradient drops}
         \label{fig:hd1}
     \end{subfigure}
     \hfill
     \begin{subfigure}[b]{0.32\textwidth}
         \centering         \includegraphics[width=0.95\textwidth]{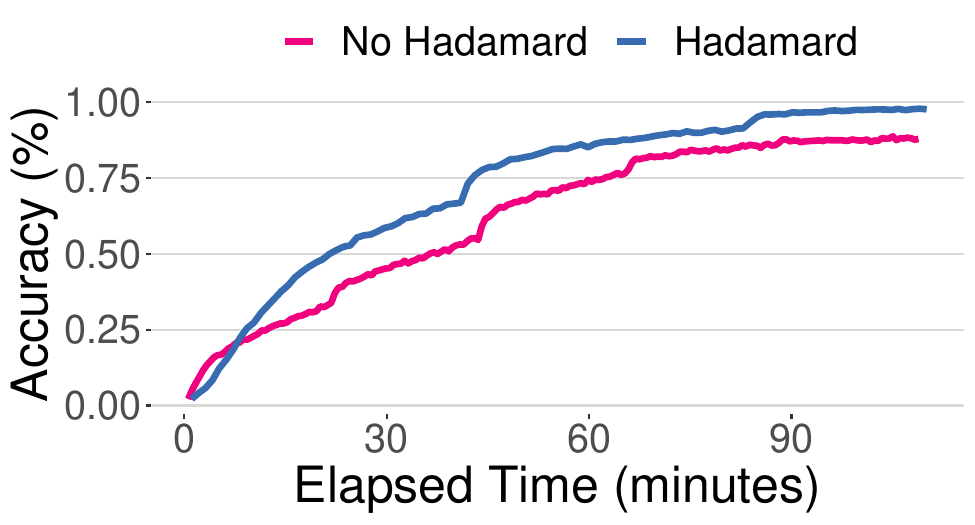}
         \vspace{-5pt}
         \caption{5\% gradient drops}
         \label{fig:hd5}
     \end{subfigure}
     \hfill
     \begin{subfigure}[b]{0.32\textwidth}
         \centering         \includegraphics[width=0.95\textwidth]{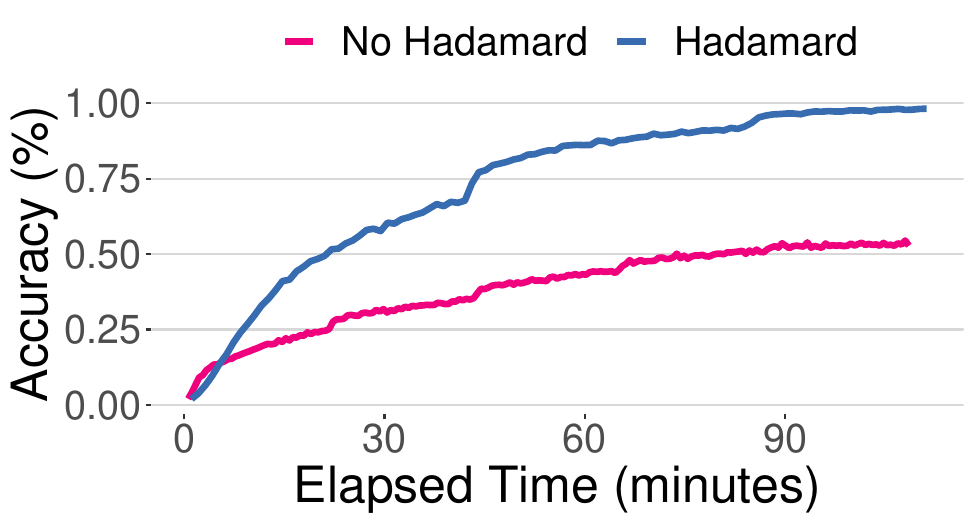}
         \vspace{-5pt}
         \caption{10\% gradient drops}
         \label{fig:hd10}
     \end{subfigure}
	     \vspace{-8pt}
        \caption{\bf Training accuracy of VGG-19 with/without Hadamard Transform (HT) in \name{}.} 
        \vspace{-15pt}
        \label{fig:hdbreak}
\end{figure*}

\begin{figure}[t]
    \centering
    \begin{subfigure}[b]{0.54\linewidth}
        \centering
        \includegraphics[width=\textwidth]{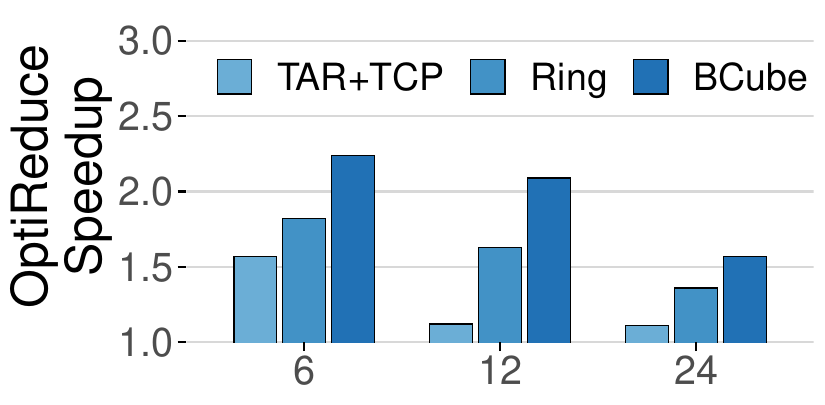}
        \vspace{-20pt}
        \caption{Local Cluster: $\bm{P_{99/50} = 1.5}$}
        \label{fig:scale-0-1}
    \end{subfigure}
    \hfill
    \begin{subfigure}[b]{0.44\linewidth}
        \centering
        \includegraphics[width=\textwidth]{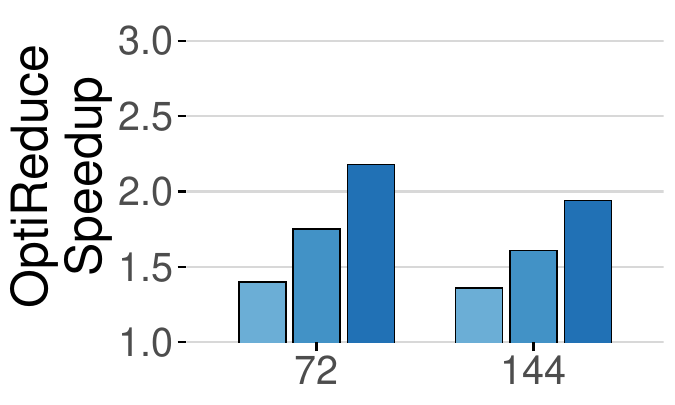}
        \vspace{-20pt}
        \caption{Sim: $\bm{P_{99/50} = 1.5}$}
        \label{fig:scale-0-1}
    \end{subfigure}
    \hfill
    \begin{subfigure}[b]{0.54\linewidth}
        \centering
        \includegraphics[width=\textwidth]{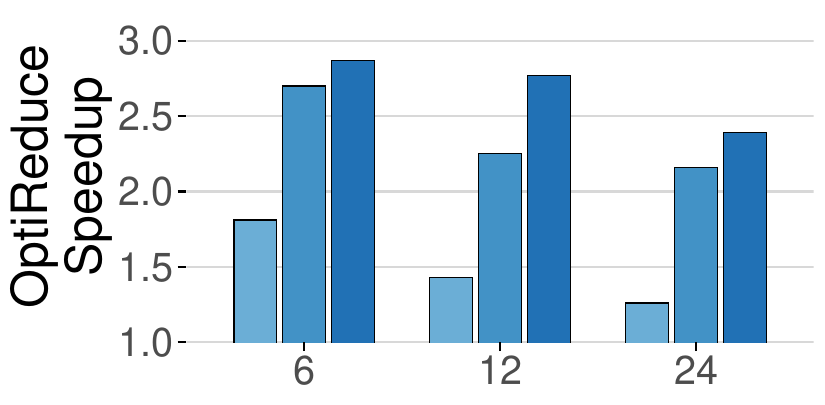}
        \vspace{-20pt}
        \caption{Local Cluster: $\bm{P_{99/50} = 3}$}
        \label{fig:scale-1}
    \end{subfigure}
    \hfill
    \begin{subfigure}[b]{0.44\linewidth}
        \centering
        \includegraphics[width=\textwidth]{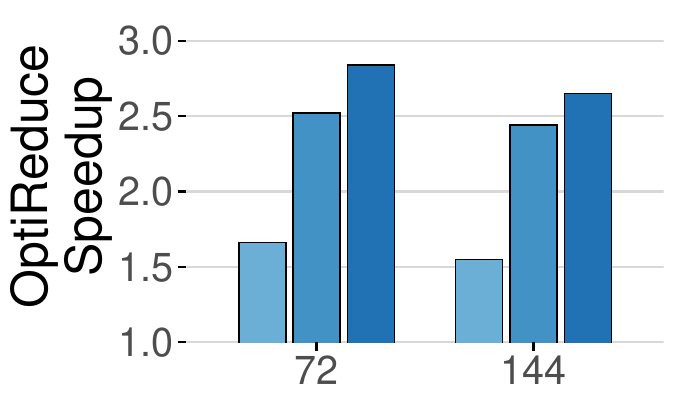}
        \vspace{-20pt}
        \caption{Sim: $\bm{P_{99/50} = 3}$}
        \label{fig:scale-0-1}
    \end{subfigure}
    \vspace{-8pt}
    \caption{\bf \name{} speedup over baseline systems (TAR+TCP, Ring, BCube) with increasing \#workers using a synthetic \SI{500}{M}-gradient AllReduce workload.} 
    \vspace{-15pt}
    \label{fig:scale}
\end{figure} 

\vspace{-3pt}
\paragraph{$\bullet$ UBT's dynamic incast improves \name{}'s latency without overloading the receiver nodes.}
We measure the effects of UBT's dynamic incast feature on \name{}'s training latency.
\Cref{fig:incast} compares two configurations: one where we fix $I = 1$, and the other with dynamically managed incast.
The results show that \name{}'s senders can leverage buffer occupancy at receivers to increase $I$, thus reducing average latency by about 21\% compared to always sending to a single receiver.
This ability to dynamically control the incast parameter ($I$) allows \name{} to adapt itself based on the capacity of the receivers' resources, which is not the case with PS (all workers send to parameter server) or Ring-AllReduce (a receiver interacts with a single sender).

\vspace{-3pt}
\paragraph{$\bullet$ \name{}'s early timeout strategy enables faster progress towards TTA.}
We evaluate the effectiveness of the early timeout strategy ($t_C$) in \name{}.
We disable $t_C$ and only keep the timeout value $t_B$ and measure its effect on training accuracy, time, and dropped gradients.
We find that when training VGG-19 with $P_{99/50} = 1.5$, \name{} takes 130 minutes to reach convergence accuracy in 200 epochs while incurring 0.02\% of gradient drops. 
Enabling early timeout brings this training time down by about 16\% (to 112 minutes) with a similar drop rate (0.02\%).
By adapting $t_C$, \name{} sustains the same drop rate and finishes quickly, rather than waiting for the higher $t_B$ value each time.
We notice that with early timeout enabled, \name{} triggers $t_C$ 95\% more often than $t_B$; hence, resulting in faster TTAs.

\vspace{-3pt}
\paragraph{$\bullet$ \name{}'s Hadamard Transform (HT) allows it to reach convergence accuracies even under higher percentages of dropped gradients.}
\Cref{fig:hdbreak} shows the training accuracy of VGG-19 model with and without Hadamard enabled. 
When considering TTA, we see that Hadamard introduces some computational overhead when operating with only 1\% of dropped gradient entries 
(\Cref{fig:hdbreak}a).
It takes Hadamard more time to reach convergence accuracies (around 97 minutes) compared to when it is disabled (90 minutes).
However, as drops increase (5\% or more), it starts to outperform the non-Hadamard instance with much faster TTAs (\Cref{fig:hdbreak}b,c).
Looking closely, we notice that across all dropped percentages, Hadamard is able to sustain the same TTA ($\approx 97$ minutes)---showing its resilience to drops. 
In contrast, the non-Hadamard case quickly degrades and fails to achieve convergence accuracy even under 10\% drops.  
The percentage drops include both drops incurred due to network variabilities (\eg, congestion and retransmissions) and gradients that a slow worker could not send due to timeouts.

\vspace{-3pt}
\paragraph{$\bullet$ \name{} scales with increasing number of worker nodes, consistently maintaining high speedups.}
To demonstrate \name{}'s performance at scale, we first run tests using CPU-based worker nodes on our local cluster. 
We compare \name{} with TAR+TCP and Gloo (Ring and BCube) on a synthetic AllReduce workload, aggregating \SI{500}{M} gradient entries across 6--24 nodes (\Cref{fig:scale}a, c).\footnote{We exclude NCCL from this comparison as it relies on GPUs.}
Next, we conduct simulations with larger clusters (72 and 144 nodes), similar in sizes to prior works~\cite{wang2024towards,lei2024seer,agarwal2024harmony,li2024flow}---using latencies sampled from the local cluster and scaled for higher node counts (\Cref{fig:scale}b, d).
Across all tests, \name{} consistently delivers high speedups, achieving 2$\times$ improvements over Ring and BCube in high-tail environments, $P_{99/50} = 3$ (\Cref{fig:scale}c, d).

\begin{figure}[t]
	\centering
	\includegraphics[width=0.78\linewidth]{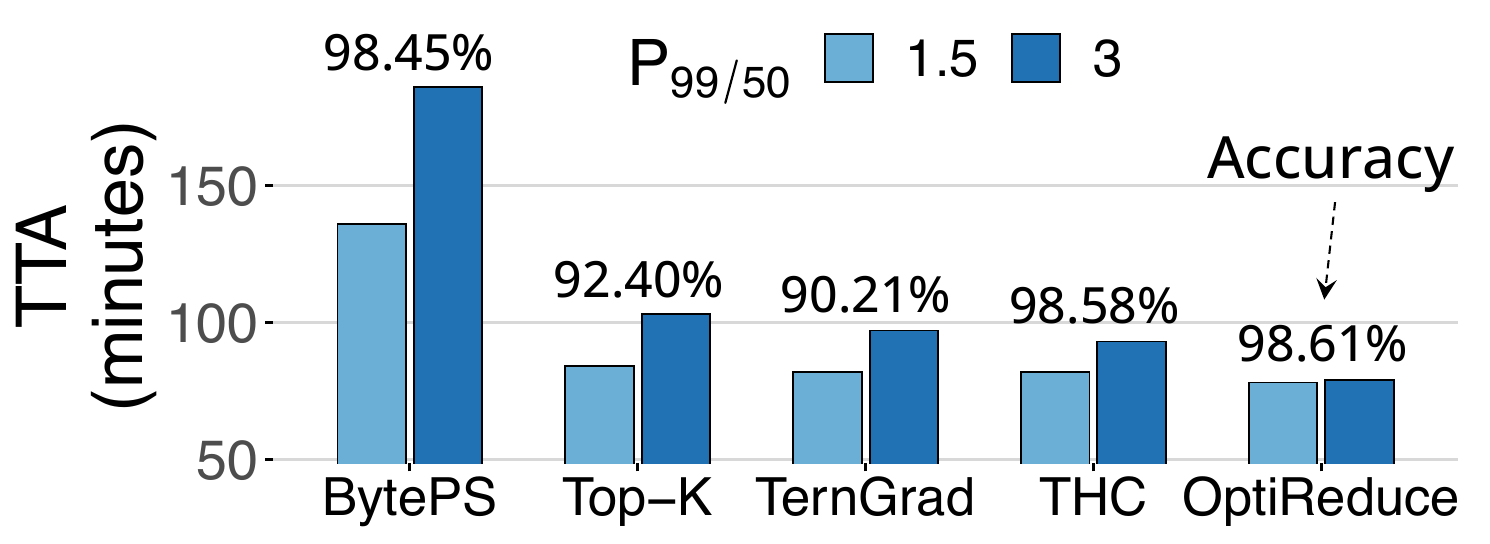}
    \vspace{-8pt}
    \caption{\bf \name{} comparison with lossy/compression schemes (BytePS, Top-K, TernGrad, \& THC), showing TTA and their convergence accuracy.} 
    \vspace{-15pt}
    \label{fig:compression}
\end{figure}

\vspace{-3pt}
\paragraph{$\bullet$ Unlike \name{}, lossy/compression schemes are vulnerable to tail effects in shared environments.}
Though \name{} is orthogonal to sparsification and quantization techniques, our comparison (\Cref{fig:compression}) shows that lossy/compression schemes (\eg, Top-K, TernGrad, and THC) fail to effectively address tail effects. 
While these schemes reduce the volume of gradient entries shared, they rely on a static evaluation of how much data to compress (or drop) a priori before transmission. 
In contrast, \name{} handles loss in real-time, dynamically adapting to network conditions and minimizing tail latency.
For instance, THC matches \name{} in convergence accuracy but takes 4\% and 18\% longer to complete under $P_{99/50} = 1.5$ and 3, respectively. 
Other schemes perform even worse, either requiring 2$\times$ more time to converge or stalling at lower accuracies due to their lossy compression, failing to improve end-to-end TTA even with additional training epochs~\cite{wang2024towards,li2024thc}.

\paragraph{$\bullet$ In-network aggregation (INA) approaches struggle with tail effects, while \name{} remains unaffected.}
In-network aggregation (INA) methods, such as SwitchML~\cite{sapio2021scaling}, reduce network latency through hardware-accelerated aggregation within the network. 
However, they remain vulnerable to tail effects---significantly inflating their completion times as the tail-to-median ratio increases.
For instance, in a low-tail environment ($P_{99/50} = 1.5$), SwitchML performs 52\% faster than \name{}. 
However, as the tail-to-median ratio increases from $P_{99/50} = 1.5$ to $3$, its completion time rises by about 2.1$\times$, even surpassing \name{} by 28\%. 
In contrast, \name{} remains unaffected by this change while reaching convergence.
By bypassing stragglers and proceeding without waiting for all gradients, \name{} is better suited for shared and high-tail environments. 
Moreover, \name{}'s design can be extended to incorporate INA, potentially achieving similar speedups in low-tail environments---an avenue we plan to explore in future work.

\vspace{-2pt}
\section{Limitations \& Future Work}
\label{sec:discussion}

In the current AllReduce design, there are two primary bottlenecks: (a) in the computation (or reduction) phase and (b) in the communication phase. 
We explore potential solutions for both bottlenecks in the subsequent sections.

\vspace{-3pt}
\paragraph{a) Accelerators for Reduction.}
In \name{}, we primarily focus on bounding the execution time of the two send/receive stages in AllReduce (\Cref{fig:epochs}).
The reduction stage, \ie, the process of averaging gradients together, still happens on CPUs. 
However, as models grow and gradient sizes increase, the reduction stage can become a bottleneck.
Rather than opting for the most extreme case of offloading all of AllReduce to network switches~\cite{sapio2021scaling, lao2021atp}, we can instead consider offloading the reduction operation on the end-host server (similar to how we accelerate GEMMs using GPUs)~\cite{steinkraus2005using}.
Modern SmartNICs~\cite{nvidianic,intelnic,napanic}, with onboard FPGAs and ML accelerators, can present a promising opportunity for accelerating reduction.
But, doing so requires rethinking and redesigning the application interface (API) between \name{} and SmartNICs.
We hope \name{} to serve as a stepping stone for research in this direction.

\vspace{-3pt}
\paragraph{b) Accelerators for Network Transport.}
As with reduction, network transport can also become a bottleneck with link/interface speeds reaching \SI{400}{Gbps+}.
Existing offloads, like RDMA~\cite{guo2016rdma}, provide high-bandwidth communication between nodes by moving data to/from the main memory and the network without engaging the host CPU. 
However, these implementations are still susceptible to tail effects in the network (\eg, packet drops and retransmissions).
We hope \name{}'s transport design can offer guidance in building new offloads for network transport, particularly with support for unreliable bounded protocols.
As a next step, we could explore offloading \name{}'s transport onto hardware using RDMA's Unreliable Connected (UC) or Unreliable Datagram (UD) queue pairs~\cite{barak2015rdma}.
However, these implementations currently suffer from excessive packet drops when packets arrive out of order~\cite{mellanoxooo}. We plan to address these challenges in future work.
\vspace{-2pt}
\section{Related Work}
\label{sec:relwork}

\vspace{-3pt}
\paragraph{Lossy Architectures for Accelerating Allreduce Collectives.} 
THC~\cite{li2024thc} presents compression-aware gradient synchronization architectures for DNN training, introducing homomorphic compression to reduce bandwidth through quantization.
OmniReduce~\cite{fei2021efficient} introduces the concept of a streaming aggregation, which exploits parameter sparsity to maximize effective bandwidth use by sending only non-zero data blocks.
MLT~\cite{wang2024towards} configures network switches to prioritize and drop packets based on model layers and gradient magnitudes, leveraging inter-packet order-independency to balance load.
In contrast, \name{} exploits DDL's resiliency to gradient drops in mitigating tail effects while sustaining convergence accuracies in the cloud without requiring access to the provider's network. 
It could apply techniques like OmniReduce to reduce network usage for models with sparse gradients or use quantization methods similar to THC.

\vspace{-3pt}
\paragraph{Accelerating Deep Learning using In-Network Computing.}
SHArP~\cite{graham2016scalable} is a hierarchical aggregation protocol and architecture in Nvidia Switches (\eg, IB-2~\cite{ib2}), which builds an overlay reduction tree for aggregating data flowing through the switch. 
SwitchML~\cite{sapio2021scaling} accelerates distributed training by using a programmable data-plane device (\eg, Intel Tofino~\cite{tofino,bosshart2013forwarding}) to aggregate the model updates from multiple workers in the network. 
To overcome the switch memory and computational constraints, they co-design the in-switch processing with end-host protocols (\eg, sliding window of parameters) for handling drops. 
ATP~\cite{lao2021atp} is an in-network aggregation solution similar to SwitchML, but for deep learning, and is designed
to provide a dynamic, best-effort in-network aggregation service for multi-tenant multi-switch clusters. 
Unlike these solutions, \name{} does not require specialized hardware or access to the provider's network.

\vspace{-3pt}
\paragraph{Optimizing Deep Learning for Enterprise and HPC Environments.}
Cassini~\cite{rajasekaran2024cassini} is a network-aware job scheduler for ML clusters in HPC environments that optimizes network resource usage by interleaving communication patterns of ML jobs, reducing congestion and improving cluster performance.
Meta's recent paper~\cite{gangidi2024rdma} presents a custom backend for distributed deep-learning training targeting enterprise data centers. 
It optimizes network topology, job scheduling, placement, and data transport to improve training performance, efficiency, and scalability.
On the other hand, \name{} offers a resilient and tail-optimal solution for deep-learning training in the cloud.

\vspace{-2pt}
\section{Conclusion}
\label{sec:conclusion}
\name{} leverages distributed-deep learning's (DDL) resiliency to lost gradients and achieves speedups of up to (70\%, 30\%), on average, over existing frameworks (Gloo, NCCL), in shared environments (\eg, public clouds).
\name{} implements a domain-specific Transpose Allreduce collective algorithm with unreliable bounded transport (UBT) featuring adaptive timeouts, while mitigating the impact of gradient loss using Hadamard Transform.
It delivers higher tail performance (\eg, TTA and training throughput) while preserving DDL models' convergence accuracy and limiting gradient drops to under 0.1\%.
Looking forward, we hope \name{} inspires further exploration of the tradeoff between tail performance and training accuracy in processing contemporary deep-learning models.

\label{lastpage}

\if\showacks1
    \vspace{-2pt}
\section*{Acknowledgements}
We sincerely appreciate the guidance of our shepherd, Changhoon Kim, as well as Ashwin Murthy, Roop Mukherjee, Leo Liu, Minlan Yu, Tushar Krishna, Ajay Brahmakshatriya, and the anonymous reviewers for their valuable feedback in strengthening this paper.
We also thank Roop Mukherjee, Leo Liu, Ali Imran, and Ali Aqdas for helping with the artifact and initial background studies on tail behavior in distributed training environments, conducted on Nvidia's internal shared clusters and AWS EC2 instances.
This work was supported in part by ACE, one of the seven centers in JUMP 2.0, a Semiconductor Research Corporation (SRC) program sponsored by DARPA; by NSF awards CAREER-2338034 and CNS-2211381; and through a Google Research Scholar Award.
Support also came in part via a generous gift from Nvidia.

\fi

{\footnotesize
\bibliographystyle{plain}
\bibliography{paper}}

\begin{figure*}[t]
  \centering
  \includegraphics[width=0.95\linewidth]{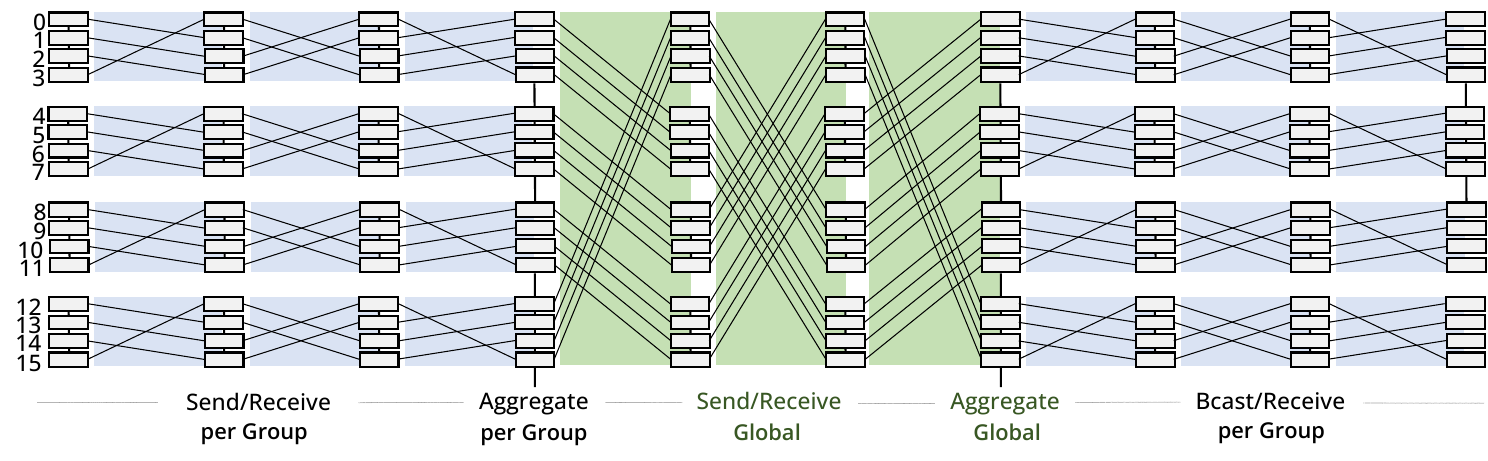}
  \vspace{-10pt}
  \caption{\bf Hierarchical 2D TAR Algorithm.}
  \label{fig:2d-tar}
  \vspace{2pt}
\end{figure*}

\begin{table*}
\begin{center}
    \small
    \begin{tabular}{l|c|cc|cc|c|>{\columncolor[HTML]{cfe2f3}}ccc}
        \toprule
        \multirow{2}{*}{{\bf Environment}} 
        & \multirow{2}{*}{{\bf Benchmark}} 
        & \multicolumn{2}{c|}{{\bf Gloo}} 
        & \multicolumn{2}{c|}{{\bf NCCL}} 
        & \multirow{2}{*}{{\bf TAR+TCP}} 
        & \multicolumn{3}{c}{\cellcolor[HTML]{cfe2f3}{\bf \name{}:}} \\ \cline{3-6}
        & 
        & \multicolumn{1}{c|}{\textbf{\em Ring}} 
        & \multicolumn{1}{c|}{\textbf{\em BCube}} 
        & \multicolumn{1}{c|}{\textbf{\em Ring}} 
        & \multicolumn{1}{c|}{\textbf{\em Tree}} 
        & 
        & \cellcolor[HTML]{cfe2f3}{\bf Time} 
        & \cellcolor[HTML]{cfe2f3}{\bf Accuracy} 
        & \cellcolor[HTML]{cfe2f3}{\bf Test Acc.} \\ \toprule

        \multirow{3}{*}{$P_{99/50} = 1.5$} 
        & ARC   & 84 & 113 & 77 & 75 & 76 & 61  & \cellcolor[HTML]{cfe2f3} 60.45 [+0.45] & \cellcolor[HTML]{cfe2f3} 39.97 [-0.47] \\ 
        & MATH  & 195 & 254 & 180 & 171 & 175 & 130 & \cellcolor[HTML]{cfe2f3} 30.56 [+0.18] & \cellcolor[HTML]{cfe2f3} 30.29 [+0.23] \\ 
        & SQuAD & 4072 & 5402 & 3391 & 3464 & 3723 & 3182  & \cellcolor[HTML]{cfe2f3} \cellcolor[HTML]{cfe2f3} 46.77 [-0.21] & \cellcolor[HTML]{cfe2f3} 38.64 [+0.08] \\ 
        \midrule

        \multirow{3}{*}{$P_{99/50} = 3.0$} 
        & ARC   & 155 & 161 & 128 & 120 & 86 & 61  & \cellcolor[HTML]{cfe2f3} 60.44 [+0.44] & \cellcolor[HTML]{cfe2f3} 39.91 [-0.53] \\ 
        & MATH  & 308 & 390 & 299 & 243 & 189 & 131 & \cellcolor[HTML]{cfe2f3} 30.14 [-0.24] & \cellcolor[HTML]{cfe2f3} 30.09 [+0.03] \\ 
        & SQuAD & 5793 & 8057 & 5677 & 5243 & 4120 & 3220  & \cellcolor[HTML]{cfe2f3} 46.45 [-0.53] & \cellcolor[HTML]{cfe2f3} 38.57 [+0.01] \\ 
        \bottomrule
    \end{tabular}
    \vspace{-8pt}
    \caption{\bf Comparing convergence time (in minutes) and accuracy (\% [$\Delta$]), as well as test accuracy (\% [$\Delta$]) for the Llama-3.2 1B model across various tasks and environments; [$\Delta$] reports deviation from the baseline accuracy (\eg, Gloo and NCCL).}
	\vspace{-18pt}
    \label{tab:llama-overview}
\end{center}
\end{table*}

\appendix

\vspace{-2pt}
\section{Hierarchical 2D TAR Algorithm: Scaling to Larger Node Clusters}
\label{appendix:hiertar}

In the hierarchical TAR design (\Cref{fig:2d-tar}), nodes are grouped to optimize both intra-group and inter-group communication, reducing the total number of rounds and connections required for AllReduce. 
For example, with $N = 64$ total nodes divided into $G = 16$ groups, each node communicates only with its corresponding rank across the groups. 
The number of rounds reduces from $2(N-1) = 126$ in traditional TAR to $2(N/G - 1) + (G - 1) = 21$ rounds.
The algorithm works in three steps:
\begin{itemize}[leftmargin=*]
	\item {\bf Intra-group Communication:} Nodes within each group perform send/receive operations followed by aggregation, in parallel, resulting in the {\em locally} aggregated shard for their rank---taking $(N/G - 1)$ rounds.
	\item {\bf Inter-group Communication:} Corresponding ranks across groups then perform send/receive operations, followed by aggregation, to get the {\em globally} aggregated shard for their rank---adding another $(G - 1)$ rounds.
	\item {\bf Broadcast Phase:} Finally, nodes within the group broadcast their aggregated shards, which are concatenated to form the globally aggregated gradient bucket---an additional $(N/G - 1)$ rounds.
\end{itemize}

This hierarchical design significantly reduces communication overhead, improving scalability and efficiency for large-scale distributed training.

\vspace{-2pt}
\section{Benchmarking Llama-3.2 1B Model}
\label{appendix:llama}

Using our local testbed (\S\ref{ssec:exp-setup}), we evaluate \name{} with the Llama-3.2 1B model~\cite{dubey2024llama} on three well-known downstream tasks: SQuAD (extractive question answering)\cite{rajpurkar2016squad}, ARC (science reasoning)\cite{clark2018think}, and MATH (symbolic mathematics)~\cite{hendrycks2021measuring}, across both low-tail ($P_{99/50} = 1.5$) and high-tail ($P_{99/50} = 3.0$) environments.
\Cref{tab:llama-overview} provides a detailed comparison of training times across all schemes. \name{} consistently demonstrates performance improvements across all tasks. 
Compared to NCCL, it achieves speedups of 1.35$\times$ on MATH, 1.25$\times$ on ARC, and 1.08$\times$ on SQuAD, averaging a 1.24$\times$ improvement. 
The gains are even more pronounced against Gloo, with speedups of 1.73$\times$, 1.61$\times$, and 1.49$\times$ respectively, averaging 1.61$\times$.
These improvements scale further under high-tail conditions, reaching speedups of up to 2.1$\times$ while preserving baseline model convergence and test accuracies.

\begin{figure*}[t]
     \centering
     \includegraphics[width=0.65\textwidth]{figures/local-legend.pdf}
     \vspace{-3pt}
     
     \begin{subfigure}[b]{0.32\textwidth}
         \centering
         \includegraphics[width=\textwidth]{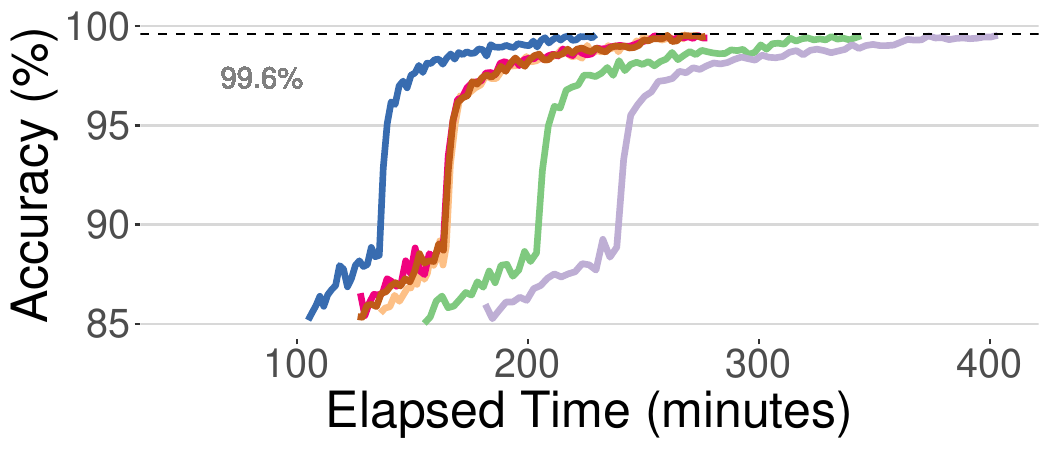}
         \vspace{-18pt}
         \caption{VGG-16}
         \label{fig:net-vgg16-0-1}
     \end{subfigure}
     \hfill
     \begin{subfigure}[b]{0.32\textwidth}
         \centering
         \includegraphics[width=\textwidth]{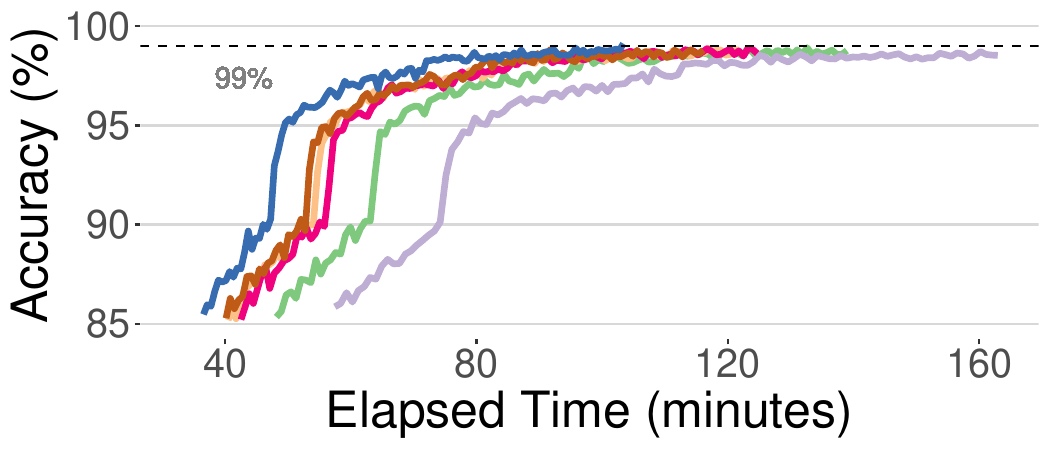}
         \vspace{-18pt}
         \caption{VGG-19}
         \label{fig:net-vgg-0-1}
     \end{subfigure}
     \hfill
     \begin{subfigure}[b]{0.32\textwidth}
         \centering
         \includegraphics[width=\textwidth]{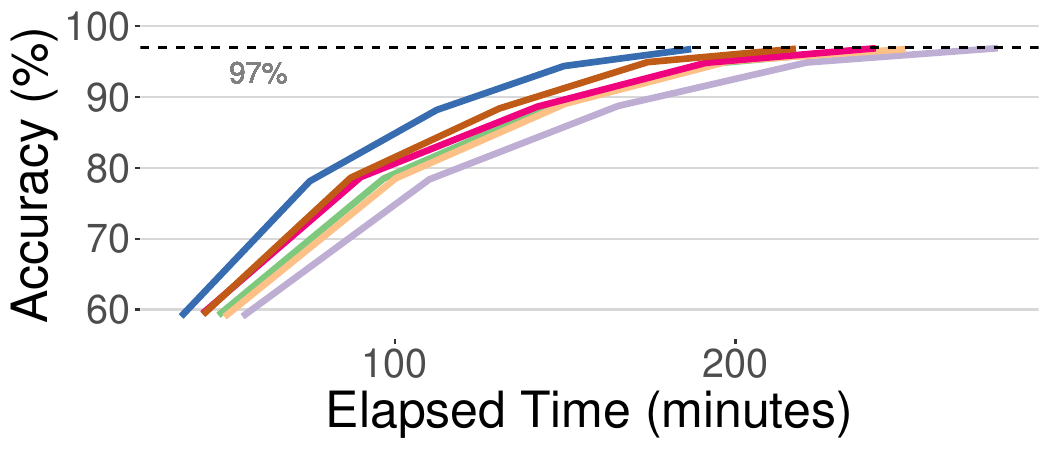}
         \vspace{-18pt}
         \caption{BERT}
         \label{fig:net-bert-0-1}
     \end{subfigure}
     \hfill
     \begin{subfigure}[b]{0.32\textwidth}
         \centering
         \includegraphics[width=\textwidth]{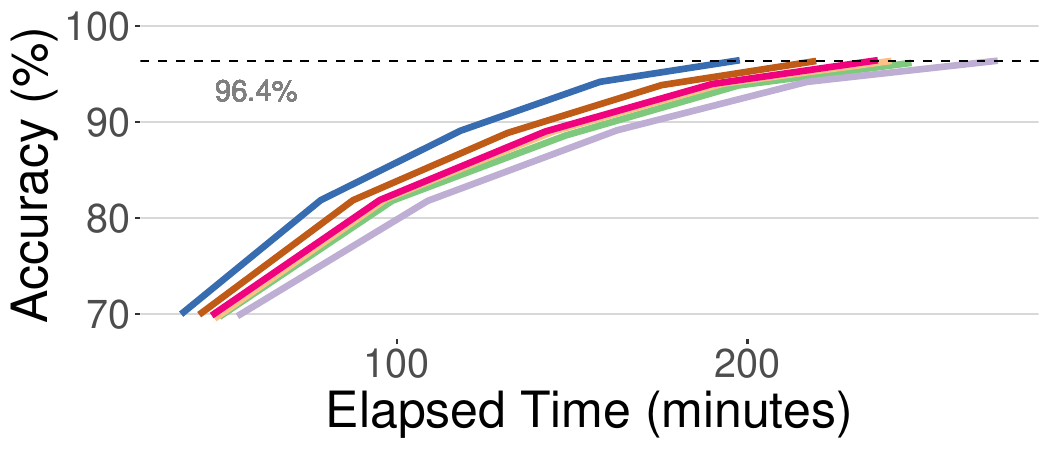}
         \vspace{-18pt}
         \caption{RoBERTa}
         \label{fig:net-roberta-0-1}
     \end{subfigure}
     \hfill
     \begin{subfigure}[b]{0.32\textwidth}
         \centering
         \includegraphics[width=\textwidth]{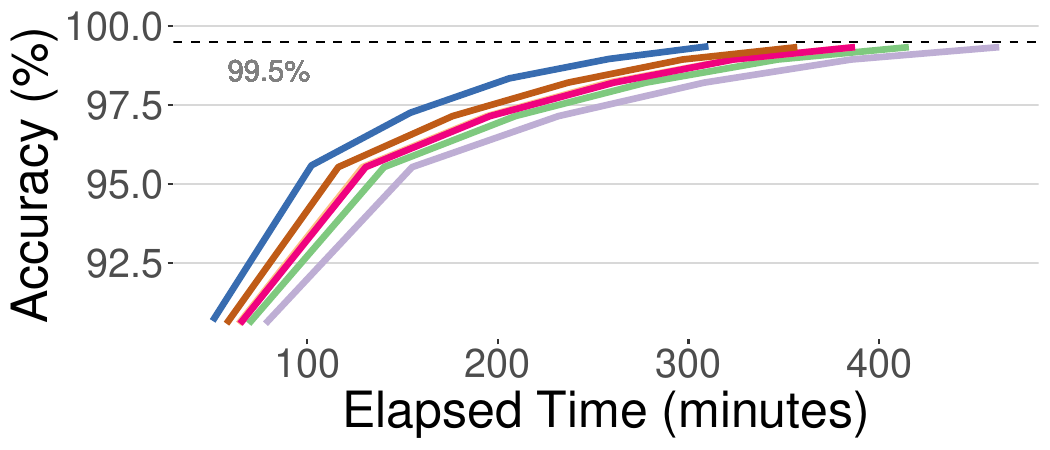}
         \vspace{-18pt}
         \caption{BART}
         \label{fig:net-bart-0-1}
     \end{subfigure}
     \hfill
     \begin{subfigure}[b]{0.32\textwidth}
         \centering
         \includegraphics[width=\textwidth]{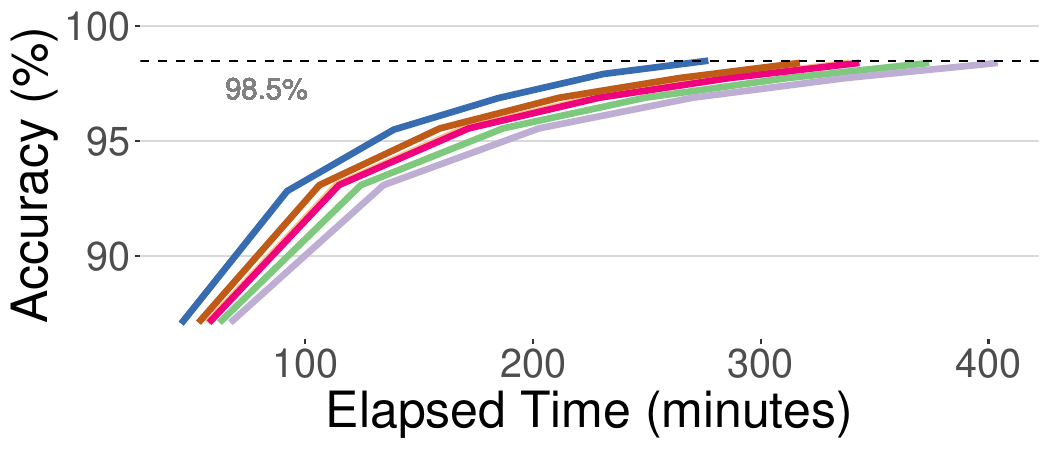}
         \vspace{-18pt}
         \caption{GPT-2}
         \label{fig:net-gpt2-0-1}
     \end{subfigure}
     
     \vspace{-8pt} 
     \caption{\bf Time-to-accuracy (TTA) of baseline systems vs \name{} with tail-to-median ratio: $\bm{P_{99/50} = 1.5}$.}
     \label{fig:net-0-1}
     \vspace{-5pt}
\end{figure*}

\begin{figure*}[t]
    \centering
    \includegraphics[width=0.65\textwidth]{figures/local-legend.pdf}
    \vspace{-3pt}

    \begin{subfigure}[b]{0.32\textwidth}
        \centering
        \includegraphics[width=\textwidth]{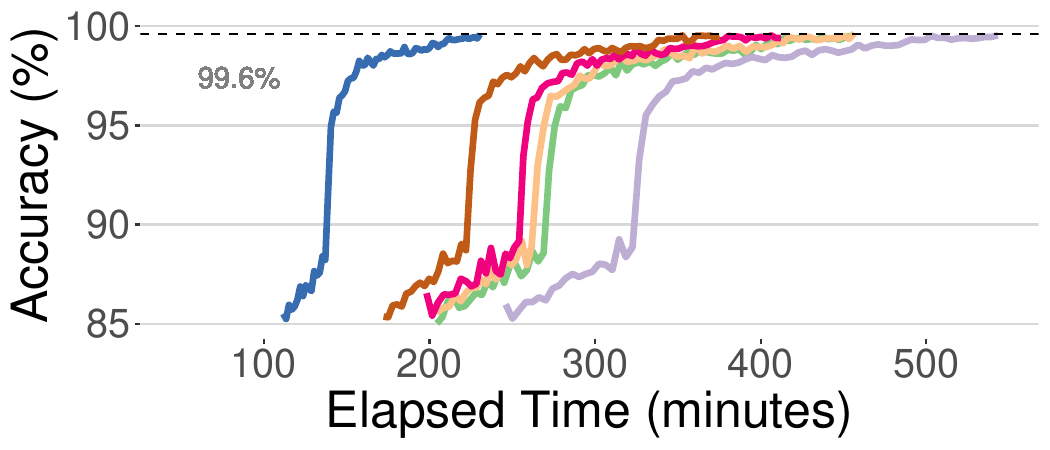}
        \vspace{-18pt}
        \caption{VGG-16}
        \label{fig:net-vgg16-1}
    \end{subfigure}
    \hfill
    \begin{subfigure}[b]{0.32\textwidth}
        \centering
        \includegraphics[width=\textwidth]{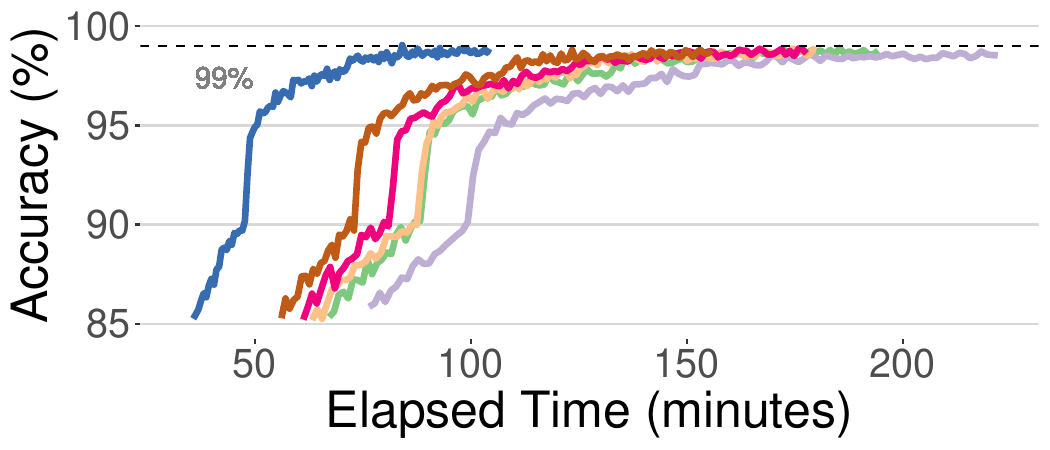}
        \vspace{-18pt}
        \caption{VGG-19}
        \label{fig:net-vgg-1}
    \end{subfigure}
    \hfill
    \begin{subfigure}[b]{0.32\textwidth}
        \centering
        \includegraphics[width=\textwidth]{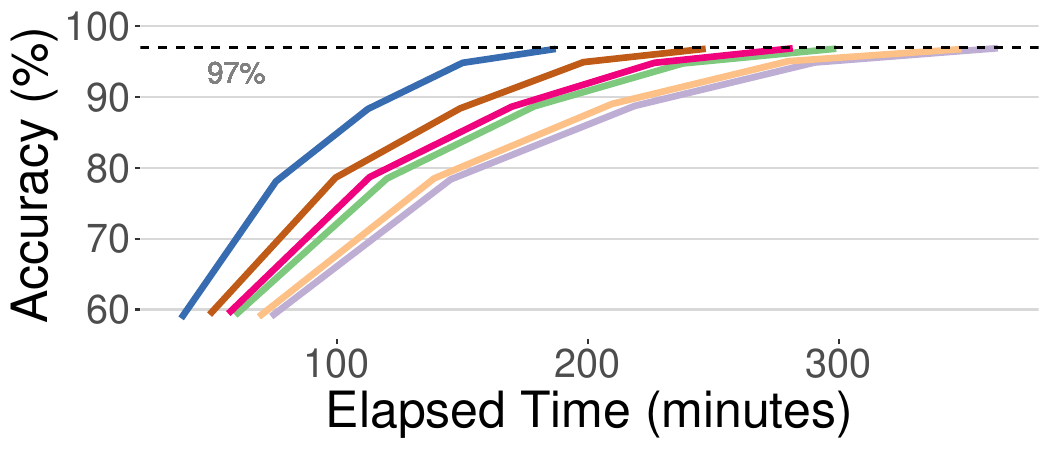}
        \vspace{-18pt}
        \caption{BERT}
        \label{fig:net-bert-1}
    \end{subfigure}

    \begin{subfigure}[b]{0.32\textwidth}
        \centering
        \includegraphics[width=\textwidth]{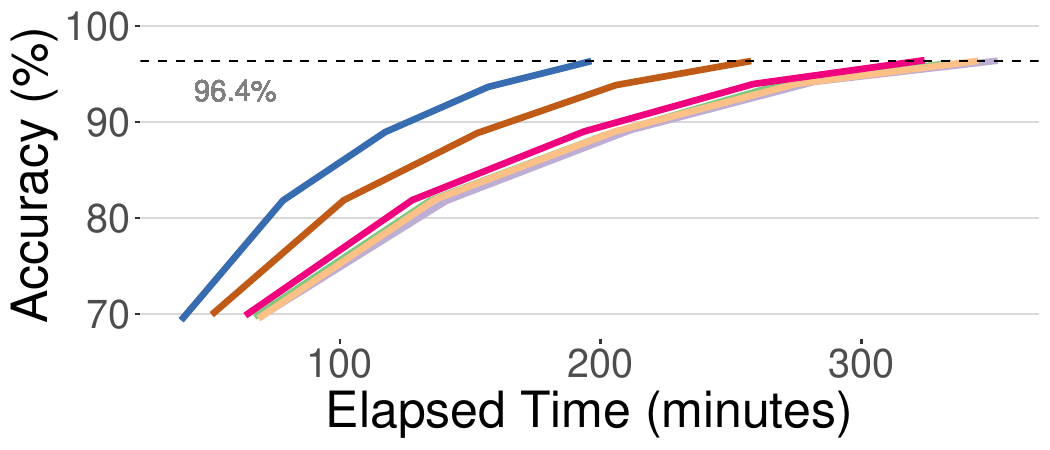}
        \vspace{-18pt}
        \caption{RoBERTa}
        \label{fig:net-roberta-1}
    \end{subfigure}
    \hfill
    \begin{subfigure}[b]{0.32\textwidth}
        \centering
        \includegraphics[width=\textwidth]{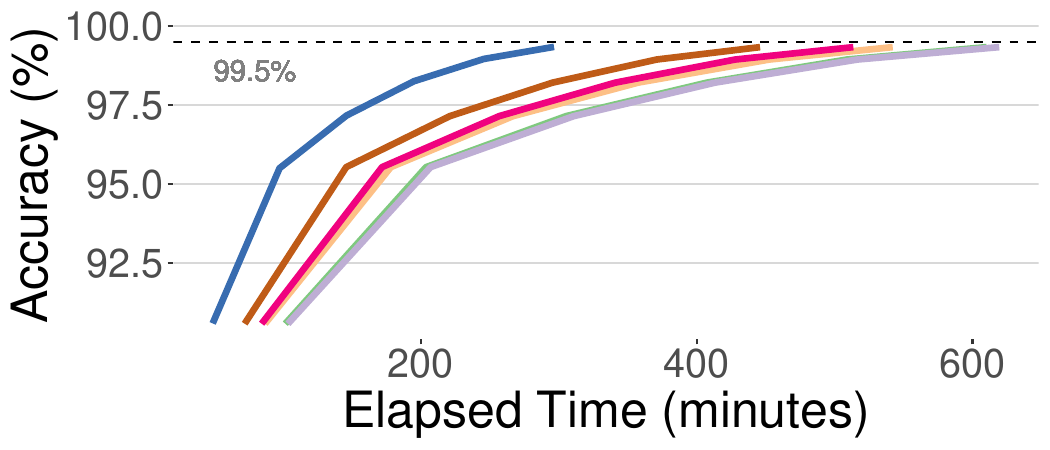}
        \vspace{-18pt}
        \caption{BART}
        \label{fig:net-bart-1}
    \end{subfigure}
    \hfill
    \begin{subfigure}[b]{0.32\textwidth}
        \centering
        \includegraphics[width=\textwidth]{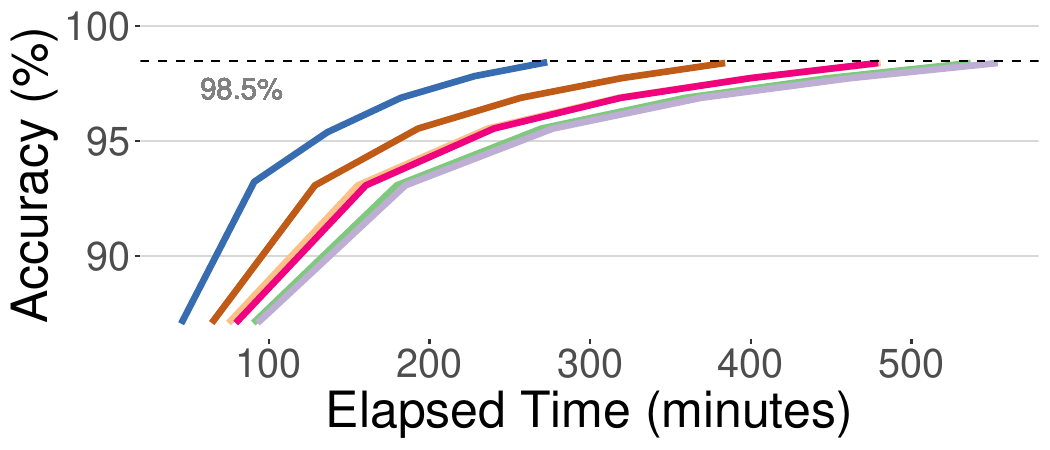}
        \vspace{-18pt}
        \caption{GPT-2}
        \label{fig:net-gpt2-1}
    \end{subfigure}

    \vspace{-8pt} 
    \caption{\bf Time-to-accuracy (TTA) of baseline systems vs \name{} with tail-to-median ratio: $\bm{P_{99/50} = 3}$.}
    \label{fig:net-1}
    \vspace{-8pt} 
\end{figure*}

\vspace{-2pt}
\section{Network and Compute Intensive Models \& Base LMs}
\label{appendix:int-llms}

In this section, we present time-to-accuracy (TTA) plots for additional models, including computer vision models (ResNet-50/101/152, VGG-16/19) and base LMs (BERT, RoBERTa, BART, and GPT-2). 
The experiments use the same local testbed setup described in \S\ref{ssec:exp-setup}, but with six worker nodes (VMs). 
We compare results across two environment configurations, characterized by tail-to-median ratios ($P_{99/50}$) of 1.5 (low variability) and 3 (high variability).

\vspace{-2pt}
\subsection{Time-to-accuracy (TTA)}
\label{appendix:tta-small}

We observe similar gains for these network-intensive models (VGG-16/19) and base LMs, with up to (66\%, 75\%) and (50\%, 51\%) reductions in TTA, on average, compared to Gloo (Ring, BCube) and NCCL (Ring, Tree), respectively---\Cref{fig:net-0-1} ($P_{99/50}=1.5$) and \Cref{fig:net-1} ($P_{99/50}=3$). 
\name{} achieves the same convergence accuracy as the baselines while limiting lost gradients to less than 1.5\%, on average, of the communicated traffic.

\vspace{-2pt}
\subsection{Training Throughput (Speedup)}
\label{appendix:resnet}

\begin{figure}[t]
 \vspace{-8pt}
 \centering  
 \includegraphics[width=0.99\linewidth]{figures/local-legend-wout-gring.pdf}
 \vspace{-2pt}
 
 \begin{subfigure}[b]{1\linewidth}
        \centering
        \includegraphics[width=0.92\linewidth]{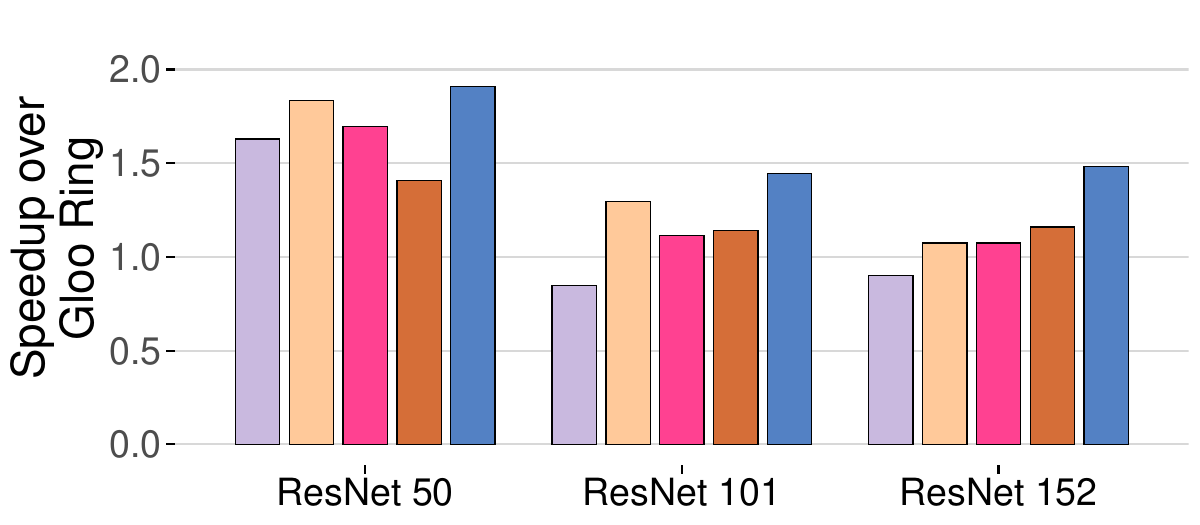}
        \vspace{-5pt}
        \caption{$\bm{P_{99/50} = 1.5}$}
        \label{fig:net-resnet-0-1}
    \end{subfigure}
    \hfill
    \begin{subfigure}[b]{1\linewidth}
        \centering
        \includegraphics[width=0.92\linewidth]{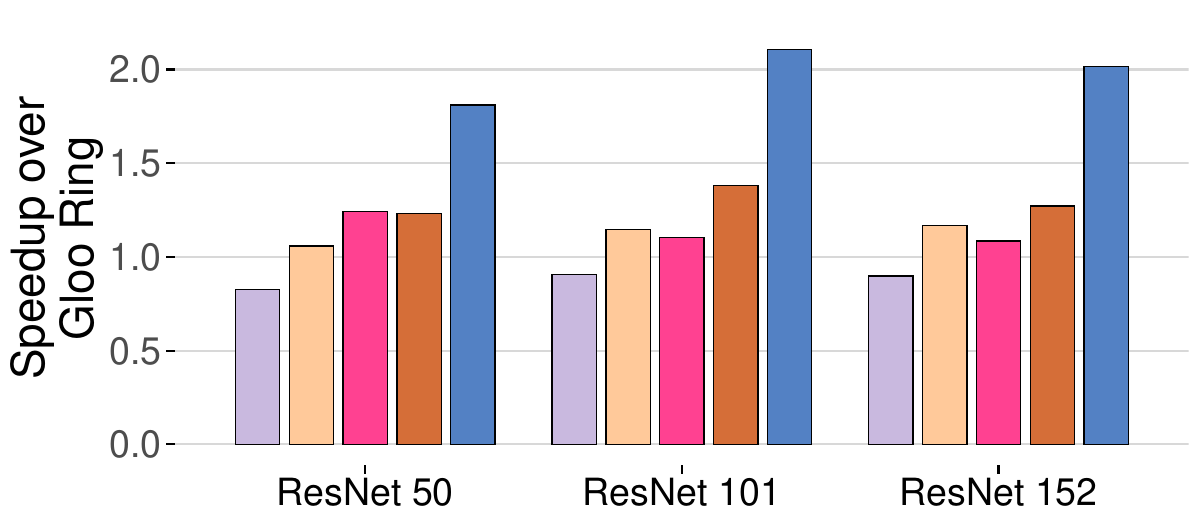}
        \vspace{-5pt}
        \caption{$\bm{P_{99/50} = 3}$}
        \label{fig:net-resnet-1}
    \end{subfigure}
\vspace{-20pt}
 \caption{\bf Training throughput for computationally-intensive ResNet models on the ImageNet dataset.}
 \label{fig:resnets}
\end{figure}

While compute-intensive models like ResNets~\cite{he2016deep} typically do not gain significant advantages from optimized communication~\cite{wang2024towards,li2024thc}, their performance can be impacted in shared environments (such as public clouds) due to long-tail latencies. 
Our evaluations reflect this, where \name{} demonstrates notable improvements over baseline systems, achieving average speedups of 22\% over NCCL and 53\% over Gloo for three ResNet models (50/101/152) across both environment configurations (\Cref{fig:resnets}).

\label{totalpage}

\end{document}